\documentclass[twocolumn]{autart}

\usepackage{cite}
\usepackage{amsmath}
\usepackage{array}
\usepackage{interval}
\usepackage{amssymb,amsfonts,mathrsfs,mathtools}
\usepackage{hyperref}
\usepackage{color}
\usepackage{booktabs,makecell}
\usepackage{pict2e}
\usepackage{flushend}
\hypersetup{colorlinks = true,
 linkcolor = red,
 urlcolor = blue,
 citecolor = blue,
 anchorcolor = blue}
\def\gof{\mbox{${\boldsymbol{P}}\,\#\,\boldsymbol{C}$}}
\newcommand\jw{j\omega}

\newcommand\bbkt[1]{\left\{#1\right\}}
\newcommand\sbkt[1]{\left[#1\right]}
\newcommand\rbkt[1]{\left(#1\right)}
\newcommand\ininf[2]{\langle #1\,, #2 \rangle}

\newcommand\lt{\mathcal{L}_2}
\newcommand\ltp{\mathcal{L}_{2}}
\newcommand\ltep{\mathcal{L}_{2e}}

\newcommand\hinf{\mathcal{H}_{\infty}}
\newcommand\rhinf{\mathcal{RH}_{\infty}}
\newcommand\ccp{\bar{\mathbb{C}}_+}
\newcommand\cn{{\mathbb{C}}^n}
\newcommand\sysp{\boldsymbol{P}}

\newcommand\sysg{\boldsymbol{G}}

\newcommand\sysi{\boldsymbol{I}}
\newcommand\sysn{\boldsymbol{N}}
\newcommand\sysc{\boldsymbol{C}}
\newcommand\sysh{\boldsymbol{H}}
\newcommand\sysm{\boldsymbol{M}}
\newcommand\sysf{\boldsymbol{F}}
\newcommand\syspu{\boldsymbol{P}u}
\newcommand\rn{{\mathbb{R}}^n}

\newcommand\abs[1]{\left|#1\right|}

\newcommand\rep{{\rm Re}}
\newcommand\imp{{\rm Im}}

\newcommand{\norm}[1]{\left\lVert#1\right\rVert}

\newcommand{\tbt}[4]{\begin{bmatrix}#1&#2\\#3&#4\end{bmatrix}}

\newcommand{\tbo}[2]{\begin{bmatrix}#1\\#2\end{bmatrix}}

\newcommand{\bi}{\begin{itemize}}\newcommand{\ei}{\end{itemize}}
\newcommand{\be}{\begin{equation}}\newcommand{\ee}{\end{equation}}
\newcommand{\bex}{\begin{equation*}}\newcommand{\eex}{\end{equation*}}
\newcommand{\bax}{\begin{align*}}\newcommand{\eax}{\end{align*}}
\newcommand{\bc}{\begin{center}}\newcommand{\ec}{\end{center}}

\newcommand{\stbt}[4]{\left[\begin{smallmatrix} #1&#2\\#3&#4\end{smallmatrix}\right]}
\newcommand{\stbo}[2]{\left[\begin{smallmatrix} #1\\#2\end{smallmatrix}\right]}

\def\goftau{\mbox{${\boldsymbol{P}}\,\#\, \rbkt{\tau\boldsymbol{C}}$}}
\intervalconfig{soft open fences}
 
\edef\endfrontmatter{
 \unexpanded\expandafter{\endfrontmatter}
 \noexpand\endNoHyper 
}

\begin{document}
\begin{frontmatter}
\title{The Singular Angle of Nonlinear Systems\thanksref{footnoteinfo}}
\thanks[footnoteinfo]{This paper was not presented at any IFAC
meeting. This work was supported in part by the European Research Council under the
Advanced ERC Grant Agreement SpikyControl n.~101054323, Research Grants
Council of Hong Kong under the General Research
Fund No.~16203922, and National Science and Technology Council of Taiwan under grants 113-2222-E-110-002-MY3 and 114-2218-E-007-011-.}
\author[CC]{Chao~Chen}\ead{chao.chen@manchester.ac.uk}, 
\author[DZ,ost]{Di~Zhao}\ead{dizhao@nju.edu.cn},
\author[SZ]{Sei~Zhen~Khong}\ead{szkhong@mail.nsysu.edu.tw}
\thanks[ost]{Corresponding author.}
\address[CC]{Department of Electrical and Electronic Engineering, The University of Manchester, Manchester M13 9PL, UK}
\address[DZ]{School of Robotics and Automation, Nanjing University, Suzhou 215163, China} 
\address[SZ]{Department of Electrical Engineering, National Sun Yat-sen University, Kaohsiung 80424, Taiwan} 

\begin{abstract}
 In this paper, we introduce an angle notion called the singular angle for nonlinear systems from an input-output perspective. The proposed system singular angle, based on the angle between $\lt$-signals, describes an upper bound for the ``rotating effect'' from system input to output signals. It quantifies passivity and serves as a counterpart to system $\lt$-gain. It also provides an alternative to a recently defined notion of system phase which adopts complexification of real-valued signals via the Hilbert transform. A nonlinear small angle theorem is established for feedback stability analysis, which involves a comparison of the loop system angle with $\pi$. The theorem generalizes the classical passivity theorem via a tradeoff between the singular angles of open-loop systems. 
\end{abstract}

\end{frontmatter}

\section{Introduction}

The story starts with a nonzero complex number $c\coloneqq \abs{c} e^{j\angle c} $ represented in the polar form. The magnitude $\abs{c}$ and argument $\angle c$, like two sides of a coin, are indispensable for $c$. Two of the most important properties of the magnitude and argument for complex numbers $a$ and $b$ are given by the identities:
\begin{align*}
 \abs{ab}=\abs{a} \abs{b}~\text{and}~\angle \rbkt{ab} =\angle a + \angle b~\mathrm{mod}~2\pi.
\end{align*}
In classical control theory, the gain (or magnitude) and phase (or angle) are two fundamental concepts for single-input single-output (SISO) linear time-invariant (LTI) systems \cite{Astrom:10}. They together form the well-known Bode diagram and are equal partners in serving control system analysis and synthesis. They have contrasting physical interpretations: the gain measures the ``amplification effect'' of physical systems while the phase measures the certain ``delay effect''. 

Over the past half-century, various efforts have been invested into generalizing the gain and phase concepts to more general systems. It is widely accepted that $\hinf$-norm \cite{Zhou:96} of multi-input multi-output (MIMO) LTI systems and $\lt$-gain \cite{Van:17} and ISS-gain \cite{Jiang:18} of nonlinear systems serve as certain roles of the gain. These gain notions all share the following crucial sub-multiplicative property:
\bex
\norm{\sysp_1\sysp_2} \leq \norm{\sysp_1} \norm{\sysp_2},
\eex
where $\|\cdot\|$ denotes a certain gain for systems $\sysp_1$ and $\sysp_2$. The nonlinear small gain theorem \cite{Zames:66} as a monumental result in the gain-based theory \cite{Van:17, Jiang:94, Hill:91} conveys a feedback stability condition involving the loop system gain being less than one. The literature on generalizations of the small gain theorem is vast and we refer the interested reader to the survey \cite{Jiang:18} for a comprehensive look. 

In contrast to the undisputed gain notion, a consensus of a proper definition about the phase counterpart is lacking among researchers, even for MIMO LTI systems. Several generations of researchers have made various attempts and efforts to search for an appropriate phase definition for MIMO LTI systems based on frequency responses. Notable works include the principal phase \cite{Postlethwaite:81}, the Bode gain/phase relation \cite{Anderson:88, Freudenberg:88, Chen:98}, the phase uncertainty \cite{Owens:84, Tits:99} and the phase margin \cite{bar-on:90}. There are also remarkable qualitatively phase-related definitions: the positive realness \cite{Anderson:73} and negative imaginariness \cite{Petersen:10, Lanzon:17, Khong:18, Lanzon:22}. Moreover, the authors of \cite{Chen:19, Chen:21} recently proposed a suitable phase definition of MIMO LTI systems and developed an LTI small phase theorem for feedback stability analysis. 

When systems become more complex, e.g., nonlinear systems, the notion of phase (or angle) is not well understood. In \cite[Sec.~6.3]{Zames:66_2}, George~Zames explicitly asked what the notion of \emph{phase-shift} is for nonlinear systems, and anticipated a stability condition involving ``\emph{loop absolute phase-shift}'' being less than $\pi$.
For a long time, passivity \cite{Van:17} has been considered as a description of phasic flavor for nonlinear systems \cite{Rantzer:19,Sepulchre:97}, since it is related to an absolute phase-shift constraint being at most $\pi/2$ in SISO LTI systems. The passivity theorem is thus treated as one realization of the anticipation in \cite[Sec.~6.3]{Zames:66_2}. Nevertheless, one may find that the passivity is only \emph{qualitatively phase-related} \cite{Chen:20j} and its connection to some ``phase constraints'' for nonlinear systems is ambiguous. Moreover, nonlinear extensions of negative imaginariness \cite{Zhao:22_NI, Ghallab:22} and counterclockwise dynamics \cite{Angeli:06} are as well qualitatively phase-related due to their SISO LTI understandings. There are also some early representative works \cite[Sec.~7.2]{Khalil:02}, \cite{Chua:79, Rugh:81, Billings:94} on investigating phasic information of nonlinear systems based on frequency-domain approximations. Recently, the authors of \cite{Chen:20j, Chen:20c} developed a phase definition for a class of sectorial nonlinear systems from an input-output perspective, and established a nonlinear small phase theorem for stability analysis as one successful realization of the aforesaid Zames' anticipation. The core idea behind the definition in \cite{Chen:20j} is to complexify real-valued signals by using the analytic signal and Hilbert transform, since the notion of phase arises naturally in a \emph{complex} domain. What if we stick to a \emph{real} domain? Can we still associate a nonlinear system with some phase (or angle) values? The answer is affirmative. 

The main purpose of this paper is to explore a brand-new angle notion, called the \emph{singular angle}, for nonlinear systems from an input-output perspective. The new notion possesses the following desirable ``additive'' property:
\bex
\theta(\sysp_1\sysp_2)\leq \theta(\sysp_1) + \theta(\sysp_2),
\eex
where $\theta(\cdot)$ denotes the singular angle of a system. This property, together with the sub-multiplicativity of the gain, has a natural association with the aforesaid crucial properties for complex numbers as a generalization. The phrase ``singular angle'', coined by Helmut~Wielandt in his lecture notes \cite[Sec.~23]{Wielandt:67}, was originally defined for complex matrices using the angle between complex vectors. Concretely, the singular angle of a square complex matrix $A\in \mathbb{C}^{n\times n}$ is defined to be 
\bex
\theta(A)\coloneqq \displaystyle \sup_{\substack{0\neq x\in \mathbb{C}^n,\\ {Ax}\neq 0}} \tilde{\theta}(x, Ax)= \sup_{\substack{0\neq x\in \mathbb{C}^n,\\ {Ax}\neq 0}} \arccos \frac{\rep \rbkt{x^* A x}}{\abs{x}\abs{Ax}},
\eex
where $|\cdot|$ represents the Euclidean norm. 
The matrix singular angle has some alternative names by other mathematicians, such as the operator angle \cite[Ch.~3]{Gustafson:97}, antieigenvalue\cite{Gustafson:94}f and operator deviation \cite{Krein:69}. It is worth noting that three of these ideas, i.e., the singular angle, operator angle, and operator deviation, were all conceived independently in different contexts in the late 1960s. We adopt the appellation ``singular angle'' since, to the best of our knowledge, \cite[Sec.~23]{Wielandt:67} is the earliest literature involving this matrix notion.

In this paper, we first adopt the angle between $\lt$-signals, which is a Hilbert space angle \cite[Ch.~3]{Gustafson:97}. Inspired by the matrix singular angle, we define the singular angle of a nonlinear system using the angles between all the input and output signals. The system singular angle quantifies the passivity; namely, the singular angle of a passive system is no greater than $\pi/2$. Meanwhile, it is related to but clearly distinct from the existing input-output passivity indices \cite{Vidyasagar:77}. In contrast to the passivity indices, the system singular angle offers an alternative approach of quantification from an angular viewpoint. Notably, it also serves as a counterpart to the system $\lt$-gain on account of the following similarities: 

\begin{enumerate}
 \renewcommand{\theenumi}{\textup{(\roman{enumi})}}\renewcommand{\labelenumi}{\theenumi}
 \item The $\lt$-gain is an operator norm induced by the $\lt$-signal norm, and the singular angle is an ``induced'' notion alike rooted in signals.
 \item The $\lt$-gain describes an upper bound for the ``stretching effect'' from system input to output signals, while the singular angle correspondingly provides an upper bound for the ``\emph{rotating effect}''. 
\end{enumerate}

A \emph{nonlinear small angle theorem} in terms of the loop system angle being less than $\pi$ is then developed for feedback stability analysis as the main result of this paper. The proposed theorem serves as a new realization of Zames' envision \cite[Sec.~6.3]{Zames:66_2}, thereby generalizing the classical passivity theorem \cite{Vidyasagar:93} via a tradeoff between the singular angles of open-loop systems. It also complements the celebrated small gain theorem \cite{Zames:66} well. The proposed theorem guarantees an ``\emph{infinite gain margin}'' of a feedback loop, and suggests a new robustness indicator of the loop, namely, a smallest ``\emph{phase margin}'' against all positive feedback gains in the loop.

The system singular angle and the recent system phase \cite{Chen:20j} are generally different with respective strengths, and are both worthy of investigation and development. The former stems from the Euclidean space angle, while the latter generalizes the phase of a complex number. The former has an advantage in studying cascaded interconnections, while the latter in investigating parallel interconnections. In short, this paper provides a new perspective of exploring the notion of phase in nonlinear systems. The angle between $\lt$-signals has been exploited in the field of control by the leading works \cite{Sontag:06, Arcak:06}, in which it is utilized to prove the secant gain stability result for the class of output strictly passive systems. By comparison, the system singular angle is defined for arbitrary stable nonlinear systems. Very recently, an incremental form of angle between $\lt$-signals was presented as a part of the scaled relative graph of nonlinear operators \cite{Ryu:21}, nonlinear systems \cite{Chaffey:21c, Chaffey:21j} and linear operators \cite{Pates:21} for convergence analysis in optimization \cite{Ryu:21} and graphical feedback stability analysis \cite{Chaffey:21j}. The scaled relative graph contains both the incremental gain and angle-type information and concentrates on graphical analysis of those graphs of systems. For comparison, this paper is dedicated to an input-output nonlinear control approach based on the brand-new singular angle. 

The remainder of this paper is structured as follows. The preliminaries on signals and systems are included in Section~\ref{sec:02}, and the singular angle of a nonlinear system is defined in Section~\ref{sec:03} equipped with crucial properties. Section~\ref{sec:04} is dedicated to a nonlinear small angle theorem for feedback stability analysis. In Section~\ref{sec:07}, we propose the frequency-wise singular angle for LTI systems for the sake of reducing conservatism. Section~\ref{sec:05} provides a link between the singular angle and passivity, and interprets the circle criterion under an ``infinite gain margin''. Section~\ref{sec:06} obtains the singular angle of a closed-loop system from that of open-loop ones. In Section~\ref{sec:09}, two variations of the singular angle are introduced and a comparison between the singular angle with the recent system phase is made. Section~\ref{sec:simulation} includes a simulation example and Section~\ref{sec:10} concludes this paper. 

\section{Notation and Preliminaries}\label{sec:02}
Let $\mathbb{F}=\mathbb{R}$ or $\mathbb{C}$ be the field of real or complex numbers, and $\mathbb{F}^n$ be the linear space of $n$-dimensional vectors over $\mathbb{F}$. Denote $\ccp$ as the closed complex right half-plane. For $x, y\in \mathbb{F}^n$, denote $\ininf{x}{y}$ and $\abs{x}\coloneqq \sqrt{\ininf{x}{x}}$ as the Euclidean inner product and norm, respectively. The conjugate, transpose and conjugate transpose of matrices are denoted by $\overline{(\cdot)}$, $(\cdot)^\top$ and $(\cdot)^*$, respectively. The real and imaginary parts of $z\in \mathbb{C}$ are denoted by $\rep\rbkt{z}$ and $\imp \rbkt{z}$, respectively. The angle of a nonzero $z\in \mathbb{C}$ in the polar form $\abs{z}e^{j\angle z}$ is denoted by $\angle z$. If $z=0$, then $\angle z$ is undefined. Denote $\rhinf^{n\times n}$ as the space consisting of $n\times n$ real rational proper matrix-valued functions with no poles in $\ccp$.

Denote by $\lt^n(-\infty, \infty)$ the set of all energy-bounded $\rn$-valued signals:~$\lt^n (-\infty, \infty) \coloneqq \{ u\colon \mathbb{R}\to \rn |~\norm{u}_2^2 \coloneqq \int_{-\infty}^{\infty} |u(t)|^2\,\mathrm{d}t < \infty\}$. The superscript $n$ is often dropped when the dimension is clear from the context. The causal subspace of $\lt(-\infty, \infty) $ is denoted by $\lt \coloneqq \{ u \in \lt(-\infty, \infty) |~u(t)=0~\text{for}~t < 0 \}$. For $T\geq 0$, define the truncation $\boldsymbol{\Gamma}_T$ on all $u\colon \mathbb{R} \rightarrow \rn$ by
$(\boldsymbol{\Gamma}_T u)(t)\coloneqq u(t)$ for $t\leq T$; $(\boldsymbol{\Gamma}_T u)(t)\coloneqq 0$ for $t>T$. 
For simplicity, we often denote $u_T \coloneqq \boldsymbol{\Gamma}_T u$ for any $T\geq 0$. Let $\ltep \coloneqq \left\{ u\colon \mathbb{R} \to \rn |~u_T \in \lt, \forall T\geq 0 \right\}$ denote the extended space of $\lt$. Let $\hat{u}$ denote the Fourier transform of a signal $u\in \lt(-\infty, \infty)$. By the well-known Plancherel's theorem, for all $u, v\in\lt(-\infty, \infty)$, we have $\ininf{u}{v} = \ininf{\hat{u}}{\hat{v}} \coloneqq \frac{1}{2\pi}\int_{-\infty}^{\infty}\hat{u}(\jw)^*\hat{v}(\jw)\, d\omega$. 

An operator $\sysp\colon \ltep\to\ltep$ is said to be causal if $\boldsymbol{\Gamma}_T\sysp=\boldsymbol{\Gamma}_T\sysp\boldsymbol{\Gamma}_T$ for all $T\geq 0$, and is said to be noncausal if it is not causal. We always assume that an operator $\sysp$ maps the zero signal to the zero signal, i.e., $\sysp0=0$. We view a system as an operator from input signals to output signals. We consider only ``square'' systems with the same number of inputs and outputs, and assume that these systems are nonzero, i.e., $\sysp\neq 0$. A nonlinear system is represented by a causal operator $\sysp\colon \ltep \rightarrow \ltep$. The $\lt$-domain of $\sysp$, namely, the set of all its input signals in $\ltp$ such that the output signals are in $\ltp$, is denoted by $\text{dom}(\sysp)\coloneqq \left\{ u\in \ltp |~ \syspu \in \ltp \right\}.$
Such a causal system $\sysp$ (operator, resp.) is said to be stable (bounded, resp.) if $\text{dom}(\sysp) = \ltp$ and
\be\label{eq:l2gain}
 \norm{\sysp} \coloneqq \sup_{0\neq u \in \ltp}  \frac{\norm{\sysp u}_2}{\norm{u}_2} < \infty.\ee
Here, $\norm{\sysp}$ is called the \emph{$\lt$-gain} of $\sysp$ and is the key quantity used in the gain-based input-output nonlinear system control theory. In addition, by \cite[Prop.~1.2.3]{Van:17}, it holds that
$\norm{\sysp} =\sup_{\substack{u \in \ltep, T>0 \\ \norm{u_T}_2\neq 0}} \frac{\norm{ (\sysp u)_T}_2}{\norm{ u_T}_2}.$
 A causal stable system $\sysp$ is called \emph{passive} {\cite{Van:17}} if
\be\label{eq:passive}
\ininf{u_T}{(\sysp u)_T}\geq 0\quad \forall u \in \ltep~\text{and}~T> 0.
\ee
Since $\sysp$ is causal and stable, it is known from \cite[Prop.~2.2.5]{Van:17} that (\ref{eq:passive}) is equivalent to
\be\label{eq:stable_passive}
\ininf{u}{\sysp u}\geq 0\quad \forall u \in \ltp.
\ee
A common practice for quantifying passivity~\eqref{eq:stable_passive} is to introduce the so-called input-output \emph{passivity indices}~\cite{Vidyasagar:77}. Specifically, a causal stable system $\sysp$ is said to be \emph{very strictly passive} if there exist $\nu, \rho>0$ such that
\be\label{eq:VSP}
\ininf{u}{\sysp u}\geq \nu \norm{u}_2^2+ \rho\norm{\sysp u}_2^2\quad \forall u \in \ltp,
\ee
where $\nu$ and $\rho$ are called the input passivity index and output passivity index, respectively. In addition, $\sysp$ is called \emph{output strictly passive} if (\ref{eq:VSP}) holds for some $\rho>0$ and $\nu=0$ for all $u \in \ltp$. Finally, $\sysp$ is called \emph{input-feedforward-output-feedback passive} if \eqref{eq:VSP} holds for certain $\nu,\rho<0$ for all $u \in \ltp$.

\section{The Singular Angle of a Nonlinear System}\label{sec:03}
This section is devoted to establishing an angle notion called the singular angle for nonlinear systems based on the angle between $\lt$-signals. It manifests that the angle between $\lt$-signals is a pseudometric function endowed with a triangle inequality. In addition, the system singular angle captures an upper bound of the ``rotating effect'' from the system input to output signals.

\subsection{The Angle Between Signals}

For $u, v\in \ltp$, we define the \emph{angle} $\theta(u,v) \in \interval{0}{\pi}$ between $u$ and $v$ by
 \be\label{eq:singular_angle}
 \theta(u, v)\coloneqq \arccos\dfrac{\ininf{u}{v}}{\norm{u}_2\norm{v}_2},\quad \text{if}~u,v \in \ltp\setminus\{0\},
 \ee
and $\theta(u, v)\coloneqq 0$, if $u=0$ or $v=0$. In light of the Cauchy-Schwarz
inequalities, i.e., $\abs{\ininf{u}{v}}\leq \norm{u}_2\norm{v}_2$ for all $u, v\in \ltp$, the ratio in (\ref{eq:singular_angle}) takes values in $\interval{-1}{1}$, and thus $\theta(u, v)$ is well defined. The angle between signals as a typical Hilbert space angle \cite[Ch.~3]{Gustafson:97} is a natural extension of the Euclidean space angle between vectors.

The following lemma introduces a useful triangle inequality of the angles between signals, which plays a significant role in feedback stability analysis in Section~\ref{sec:04}. This lemma is a modification of \cite[Lem.~3.3-1]{Gustafson:97} and the result in \cite{Krein:69}, and thus its proof is omitted.
 \begin{lem}\label{lem:signal_triangle_ineq}
 For all $u, w\in \ltp$ and $v \in \ltp\setminus\{0\}$, it holds that
 $\theta(u,w)\leq \theta(u,v)+\theta(v,w)$.
 \end{lem}

Lemma~\ref{lem:signal_triangle_ineq} indicates that the angle between signals $\theta\colon\rbkt{\ltp\setminus \{0\}} \times \rbkt{\ltp \setminus \{0\}} \rightarrow \interval{0}{\pi} $ is a pseudometric on account of the following three properties:
\begin{enumerate}
 \renewcommand{\theenumi}{\textup{(\roman{enumi})}}\renewcommand{\labelenumi}{\theenumi}
 \item pseudo identity of indiscernibles: $\theta(u,v)= 0$ if and only if $u=kv$ for some scalar $k>0$;
 \item symmetry: $\theta(u,v)=\theta(v,u)$;
 \item a triangle inequality: $\theta(u,w)\leq \theta(u,v)+\theta(v,w)$
\end{enumerate}
for all $u, v, w\in \ltp\setminus \{0\}$. Additionally, the following property also holds as expected:
$\theta(k u, v)= \theta(u, v)$ if $k>0$ and $\theta(k u, v)= \pi-\theta(u, v)$ if $k<0$.

\subsection{The Singular Angle of a Nonlinear System}
 Having introduced the angle between signals, we proceed to define the singular angle of a nonlinear system associated with the input and output signal pairs. Consider a causal stable system $\sysp:\ltep\rightarrow\ltep$. The singular angle of $\sysp$, denoted by $\theta(\sysp)\in \interval{0}{\pi}$, is defined via
 \be\label{eq:nonlinear_angle}
 \theta(\sysp)\coloneqq \displaystyle\sup_{\substack{ 0\neq u \in \ltp, {\sysp u\neq 0}}} \theta(u, \sysp u).
 \ee
Or equivalently, we note the following cosine form:
\bex
 \cos\theta(\sysp)= \inf_{\substack{ 0\neq u \in \ltp,\\{\sysp u\neq 0}}} \dfrac{\ininf{u}{\sysp u}}{\norm{u}_2\norm{\sysp u}_2}
 \eex
which will be frequently utilized in the rest of this paper.
Here, we allow a slight abuse of notation that $\theta(\cdot)$ denotes the singular angle of a system and $\theta(\cdot, \cdot)$ the angle between two signals; whenever there is no confusion.

The $\lt$-gain defined in \eqref{eq:l2gain} describes an upper bound for the ``stretching effect'' of a system from the input to output signals. Likewise, the singular angle proposed in \eqref{eq:nonlinear_angle} can be interpreted as an upper bound for the ``\emph{rotating effect}'' of a system from the input to output signals. The imagination of the ``rotating effect'' is naturally borrowed from the Euclidean space angle.

Based on the singular angle, an equivalent characterization of stable passive systems can be obtained as follows.
\begin{prop}\label{prop:angle_passive}
Let $\sysp$ be a causal stable system. Then, $\sysp$ is passive if and only if $\theta(\sysp)\in \interval{0}{\pi/2}$.
\end{prop}
\pf
By definitions~\eqref{eq:stable_passive} and \eqref{eq:nonlinear_angle}, for all nonzero $u\in \lt$ and $\sysp u \in \lt$, $\ininf{u}{\sysp u}\geq 0$ if and only if $\frac{\ininf{u}{\sysp u}}{\norm{u}_2\norm{\sysp u}_2}\geq 0$. Hence, $\sysp$ is passive if and only if $0\leq \theta(\sysp)\leq \frac{\pi}{2}$. \hfill $\square$
\endpf
Apparently the singular angle quantifies the passivity from an angular perspective. We will elaborate on more connections of the singular angle and passivity in Section~\ref{sec:05}.

The singular angle has its advantages in studying cascaded interconnections analogously to the $\lt$-gain. For given systems $\sysp_1$ and $\sysp_2$, a cascaded interconnected system $\sysp$ is defined to be $\sysp=\sysp_2\sysp_1$. Recall that the $\lt$-gain has the following sub-multiplicative property:
\bex
\norm{\sysp} \leq \norm{\sysp_1} \norm{\sysp_2}
\eex
for stable $\sysp_1$ and $\sysp_2$. The following proposition presents a parallel result on how $\theta(\sysp)$ is related, via an ``\emph{additive property}'', to $\theta(\sysp_1)$ and $\theta(\sysp_2)$.
\begin{prop}\label{prop:productPC}
 For causal stable systems $\sysp_1$ and $\sysp_2$, the cascaded interconnected system $\sysp=\sysp_2\sysp_1$ satisfies
 \bex
\theta(\sysp)\leq \theta(\sysp_1)+\theta(\sysp_2).
 \eex
\end{prop}
\pf
Let $y_1=\sysp_1 u_1$ and $y_2=\sysp_2 u_2$. Since $y_1=u_2$, for all $u_1, \sysp_1 u_1, \sysp_2\sysp_1 u_1 \in \ltp\setminus\{0\}$, it holds that
 \begin{align*}
 \theta(u_1, y_2)&=\theta(u_1, \sysp_2\sysp_1 u_1)\\
 &\leq \theta(u_1, \sysp_1 u_1)+\theta(\sysp_1 u_1, \sysp_2\sysp_1 u_1)
 \end{align*}
according to Lemma~\ref{lem:signal_triangle_ineq}. Thus, we have
 \begin{equation*}
 \begin{aligned}
 \theta(\sysp_2\sysp_1) &\leq \sup_{\substack{ 0\neq u_1 \in \ltp,\\{\sysp_1 u_1\neq 0},\\{\sysp_2 \sysp_1 u_1\neq 0}}} \sbkt{\theta(u_1, \sysp_1 u_1)+\theta(\sysp_1 u_1, \sysp_2\sysp_1 u_1)} \\
 &\leq \sup_{\substack{ 0\neq u_1 \in \ltp,\\{\sysp_1 u_1\neq 0}}} \theta(u_1, \sysp_1 u_1) +\sup_{\substack{ 0\neq u_2 \in \ltp,\\{\sysp_2 u_2\neq 0}}} \theta( u_2, \sysp_2 u_2)\\
 &= \theta(\sysp_1)+\theta(\sysp_2). \hspace{4cm}\square
 \end{aligned} 
 \end{equation*} 
\endpf

Proposition~\ref{prop:productPC} also holds for the cascaded interconnection of $N$ subsystems, namely, $\sysp\coloneqq \sysp_1\sysp_2 \cdots \sysp_N$. In this case, we will arrive at $\theta(\sysp)\leq \theta(\sysp_1) + \theta(\sysp_2)+\cdots+\theta(\sysp_N)$. This will pave the way for studying of cyclic systems in Section~\ref{sec:cyclic} via the use of singular angles.

\section{A Nonlinear Small Angle Theorem}\label{sec:04}
This section presents the main result of this paper, a nonlinear small angle theorem. The theorem states a brand-new feedback stability condition that the loop system singular angle is required to be less than $\pi$. The theorem can be regarded as an angular complement to the famous small gain theorem.

\begin{figure}[htb]
\centering
\setlength{\unitlength}{1mm}
\begin{picture}(50,25)
\thicklines \put(0,20){\vector(1,0){8}} \put(10,20){\circle{4}}
\put(12,20){\vector(1,0){8}} \put(20,15){\framebox(10,10){$\sysp$}}
\put(30,20){\line(1,0){10}} \put(40,20){\vector(0,-1){13}}
\put(38,5){\vector(-1,0){8}} \put(40,5){\circle{4}}
\put(50,5){\vector(-1,0){8}} \put(20,0){\framebox(10,10){$\sysc$}}
\put(20,5){\line(-1,0){10}} \put(10,5){\vector(0,1){13}}
\put(5,10){\makebox(5,5){$y_2$}} \put(40,10){\makebox(5,5){$y_1$}}
\put(0,20){\makebox(5,5){$e_1$}} \put(45,0){\makebox(5,5){$e_2$}}
\put(13,20){\makebox(5,5){$u_1$}} \put(32,0){\makebox(5,5){$u_2$}}
\put(10,10){\makebox(6,10){$-$}}
\end{picture}\caption{A feedback system $\gof$.} \label{fig:feedback}
\end{figure}

Consider the feedback system shown in Fig.~\ref{fig:feedback}, where $\boldsymbol{P}\colon \ltep \rightarrow \ltep$ and $\boldsymbol{C}\colon \ltep \rightarrow \ltep$ are two causal stable systems, $e_1$ and $e_2$ are external signals, and $u_1, u_2, y_1$ and $y_2$ are internal signals. Let $\gof$ denote this feedback system. Algebraically, we have the following equations:
\bex
u=e-\stbt{0}{\boldsymbol{I}}{-\boldsymbol{I}}{0}y\quad\text{and}\quad y=\stbt{\boldsymbol{P}}{0}{0}{\boldsymbol{C}}u,
\eex
where $u=\left[u_1^\top~u_2^\top\right]^\top$, $e=\left[{e_1^\top}~{e_2^\top}\right]^\top$ and $y=\left[{y_1^\top}~{y_2^\top}\right]^\top$. We assume that all the feedback systems in this paper are well-posed in the following sense. 
\begin{defn}
 A feedback system $\gof$ is said to be \emph{well-posed} if $u \mapsto e: \ltep \rightarrow \ltep = \stbt{\sysi}{\sysc}{-\sysp}{\sysi}\eqqcolon\boldsymbol{F}_{\sysp, \sysc}$ 
has a causal inverse on $\ltep$.
\end{defn}
The input-output feedback stability is defined as follows.
\begin{defn}\label{def:feedback_stability}
 A well-posed $\gof$ is said to be \emph{stable} if there exists $c>0$ such that
 $\norm{\boldsymbol{\Gamma}_T u}_2 \leq c \norm{\boldsymbol{\Gamma}_T e}_2$
 for all $T\geq 0$ and for all $e\in \ltep$, i.e., $\norm{(\boldsymbol{F}_{\sysp, \sysc})^{-1}}<\infty$.
\end{defn}

We in the following consider a special structure of $\gof$, namely, a single-loop feedback system obtained by setting the external signal $e_2=0$ in Fig.~\ref{fig:feedback}. Let $\gof|_{e_2=0}$ denote this single-loop feedback system. The stability of $\gof|_{e_2=0}$ can be defined in a similar fashion to Definition~\ref{def:feedback_stability}. The motivation is that very often it is sufficient to study $\gof|_{e_2=0}$ when we investigate nonlinear feedback systems \cite{Megretski:97}. In particular, when $\sysc$ is a linear system, the effect of $e_2$ in $\gof$ can be included in that of $e_1$ \cite[Sec.~8]{Jonsson:01}. In this case, the stability of $\gof$ is equivalent to that of $\gof|_{e_2=0}$.
 
We revisit a fundamental version of the \emph{nonlinear small gain theorem} \cite[Sec.~2.1]{Van:17}, \cite{Zames:66}: For causal stable $\sysp$ and $\sysc$, the well-posed $\gof|_{e_2=0}$ is stable if
\be \label{eq:small-gain}
\norm{\sysp}\norm{\sysc}<1.
\ee
 In what follows, we in parallel establish the so-called \emph{nonlinear small angle theorem} which ensures feedback stability using singular angles. 
\begin{thm}[Small angle theorem]\label{thm:spt}
 For causal stable $\sysp$ and $\sysc$, the well-posed $\gof|_{e_2=0}$ is stable if
\be\label{eq:spt}
\theta(\sysp) + \theta(\sysc) < \pi.
\ee
\end{thm}
\pf
When special zero signals are involved in the feedback loop, namely, $u_1=0$, $u_2=0$ or $y_2=0$, the stability of $\gof|_{e_2=0}$ can be shown separately. When $u_1=0$, it holds that $y_1=u_2=y_2=0$; When $u_2=0$, the stability of $\gof|_{e_2=0}$ is deduced from the open-loop stability of $\sysp$; When $y_2=0$, $\gof|_{e_2=0}$ becomes a cascaded open-loop system $\sysc\sysp$. Since $\sysc$ and $\sysp$ are stable, then $\sysc\sysp$ is stable. Therefore, it suffices to show the proof for the case $u_1, u_2, y_2 \in \ltp\setminus \{0\}$. The singular angle as a system property is defined on $\ltp$, while the well-posedness and closed-loop stability are defined on $\ltep$. To deal with this, we use a homotopy argument with several steps similar to that used in \cite{Megretski:97} to prove the result.

\emph{Step 1}: For all $u \in \ltp\setminus\{0\}$ and $\tau \in \interval{0}{1}$, we show that there exists $c_0>0$, independent of $\tau$, such that 
 $\norm{u}_2\leq c_0 \norm{ \sysf_{\sysp, \tau\sysc} u}_2$. 

Let $y_1=\sysp u_1$ and $y_2=\tau\sysc u_2$. When $\tau=0$, $\goftau|_{e_2=0}$ is stable by the open-loop stability. We only need to consider the case $\tau\in\interval[open left]{0}{1}$ and note that $\theta(\sysc)=\theta(\tau\sysc)$ for all $\tau\in \interval[open left]{0}{1}$. Since $\cos(\cdot)$ is a decreasing function on $\interval{0}{\pi}$, by hypothesis (\ref{eq:spt}), we have $\cos\sbkt{\theta(\sysp) + \theta(\tau\sysc)} > -1$.
The above inequality implies
 \bex
 \cos\sbkt{\theta(u_{1}, y_{1}) + \theta(u_{2}, y_{2})} > -1
 \eex
for all $u_1, u_2\in\ltp\setminus\{0\}$ and $y_1, y_2\neq 0$. By Lemma~\ref{lem:signal_triangle_ineq} and $y_1=u_2$, we have
 \bex
 -1<\cos\sbkt{\theta(u_1, y_1) + \theta(u_2, y_2)}\leq \cos\theta(u_1, y_2) \leq 1
 \eex
for all $u_1, u_2\in\ltp\setminus\{0\}$ and $y_1, y_2\neq 0$. Note that 
\begin{multline}\label{eq:sysineq}
\hspace{-3mm}\norm{ \sysf_{\sysp, \tau\sysc} u}^2_2=\norm{\tbo{u_1+y_2}{0}}^2_2 =\norm{u_1}^2_2+\norm{y_2}^2_2+2\ininf{u_1}{y_2}\\
\geq\norm{u_1}^2_2+\norm{y_2}^2_2+2\norm{u_1}_2\norm{y_2}_2\cos\theta(u_1, y_2).
\end{multline}
Firstly, we assume $\theta(u_1, y_2)\in \interval{0}{{\pi}/{2}}$, and thus $ \cos\theta(u_1, y_2)\in \interval{0}{1}$. In this case, discarding the nonnegative terms $\norm{y_2}_2$ and $2\norm{u_1}_2\norm{y_2}_2\cos\theta(u_1, y_2)$ in (\ref{eq:sysineq}) yields
\bex
\norm{u_1}_2\leq \norm{\sysf_{\sysp, \tau\sysc} u}_2. 
\eex
Since $\sysp$ is stable and $u_2=y_1$, we have
\bex
\norm{u_2}_2=\norm{\sysp u_1}_2\leq \norm{\sysp}\norm{u_1}_2\leq \norm{\sysp}\norm{\sysf_{\sysp, \tau\sysc} u}_2. 
\eex
Therefore, it suffices to show the stability for the case when $\theta(u_1, y_2)\in\interval[open]{\pi/2}{\pi}$. By using (\ref{eq:sysineq}), we obtain
\begin{align*} 
&\norm{\sysf_{\sysp, \tau\sysc} u}^2_2\geq \norm{u_1}^2_2+\norm{y_2}^2_2+2\norm{u_1}_2\norm{y_2}_2\cos\theta(u_1, y_2)\\
&=\sbkt{\cos\theta(u_1, y_2)}^2\norm{u_1}^2_2+2\norm{u_1}_2\norm{y_2}_2\cos\theta(u_1, y_2)\\
&\quad +\norm{y_2}^2_2 +\{1-\sbkt{\cos\theta(u_1, y_2)}^2\}\norm{u_1}_2^2 \\
&=\sbkt{{\cos\theta(u_1, y_2)}\norm{u_1}_2+\norm{y_2}_2}^2\\
&\quad +\{1-\sbkt{\cos\theta(u_1, y_2)}^2\}\norm{u_1}_2^2\\
&\geq \{1-\sbkt{\cos\theta(u_1, y_2)}^2\}\norm{u_1}_2^2.
\end{align*}
Since $\cos\theta(u_1, y_2)\in\interval[open]{-1}{0}$ and $\sbkt{\cos\theta(u_1, y_2)}^2\in~\interval[open]{0}{1}$, there exists a constant $c_1>0$ such that
\begin{align*}
\norm{u_1}_2&\leq \frac{1}{\sqrt{1-\sbkt{\cos\theta(u_1, y_2)}^2}}\norm{\sysf_{\sysp, \tau\sysc} u}_2 \\
&=\frac{1}{\sin\theta(u_1, y_2)} \norm{ \sysf_{\sysp, \tau\sysc} u}_2 \\
&\leq \frac{1}{\sin\sbkt{\theta(\sysp)+\theta(\tau\sysc)}}\norm{ \sysf_{\sysp, \tau\sysc} u}_2 \\
 &=\frac{1}{\sin\sbkt{\theta(\sysp)+\theta(\sysc)}}\norm{ \sysf_{\sysp, \tau\sysc} u}_2 
\eqqcolon c_1\norm{ \sysf_{\sysp, \tau\sysc} u}_2,
\end{align*}
where $\pi/2<\theta(u_1, y_2)\leq \theta(\sysp)+\theta(\tau\sysc)<\pi$ and $\theta(\tau\sysc)=\theta(\sysc)$ when $\tau\in\interval[open left]{0}{1}$. Moreover, since $\sysp$ is stable, there exists a constant $c_2\coloneqq c_1\norm{\sysp}>0$ such that
\begin{align*}
 \norm{u_2}_2=\norm{\sysp u_1}_2\leq \norm{\sysp}\norm{u_1}_2\leq
 c_2\norm{ \sysf_{\sysp, \tau\sysc} u}_2. 
\end{align*}
Note that $c_1>1$. Thus, we can unify the two cases $\theta(u_1, y_2)\in\interval{0}{{\pi}/{2}}$ and $\theta(u_1, y_2)\in\interval[open]{\pi/2}{\pi}$ into the following inequality:
\begin{align*}
 \norm{u}_2&\leq \norm{u_1}_2+\norm{u_2}_2 \leq (c_1+c_2)\norm{ \sysf_{\sysp, \tau\sysc} u}_2. 
\end{align*}
Then, there exists a constant $c_0\coloneqq c_1+c_2>0$, independent of $\tau$, such that for all $ u \in \ltp\setminus\{0\}$ and for all $\tau \in \interval{0}{1}$, we have $\norm{u}_2\leq c_0 \norm{ \sysf_{\sysp, \tau\sysc}u}_2$. 

\emph{Step 2}: Show that the stability of $\boldsymbol{P}\,\#\,\rbkt{\tau\boldsymbol{C}} |_{e_2=0}$ implies the stability of $\boldsymbol{P}\,\#\,\sbkt{(\tau+\nu)\boldsymbol{C}} |_{e_2=0}$ for all $\abs{\nu}< \mu= {1}/({c_0\norm{\boldsymbol{C}}})$, where $\mu$ is independent of $\tau$.
  
By the well-posedness assumption, the inverse $\rbkt{\sysf_{\sysp, \tau\sysc}}^{-1}$ is well-defined on $\ltep$. By hypothesis, $\rbkt{\sysf_{\sysp, \tau\sysc}}^{-1}$ is bounded on $\ltp$. Given $u\in \ltep$, we define
$u_T\coloneqq \rbkt{\sysf_{\sysp, \tau\sysc}}^{-1}\boldsymbol{\Gamma}_T\rbkt{\sysf_{\sysp, \tau\sysc}u} \in \ltp$,
where an abuse of the subscript notation $T$ in $u_T$ is allowed since there is no confusion with the truncation operator $\boldsymbol{\Gamma}_T$. Then 
\begin{align*}
 & \norm{\boldsymbol{\Gamma}_T u}_2=\norm{\boldsymbol{\Gamma}_T u_T}_2\leq \norm{u_T}_2\\
 \leq &~c_0\norm{\sysf_{\sysp, \tau\sysc} u_T}_2=c_0\norm{\boldsymbol{\Gamma}_T\rbkt{\sysf_{\sysp, \tau\sysc}u}}_2\\
 \leq &~ {c_0\norm{\boldsymbol{\Gamma}_T\rbkt{\tbt{\boldsymbol{I}}{(\tau+\nu)\boldsymbol{C}}{-\boldsymbol{P}}{\boldsymbol{I}} u}-\boldsymbol{\Gamma}_T\rbkt{\tbt{0}{\nu \boldsymbol{C}}{0}{0}u}}_2}\\
 = &~{c_0\norm{\boldsymbol{\Gamma}_T\rbkt{\sysf_{\sysp, (\tau+\nu)\sysc} u}-\boldsymbol{\Gamma}_T\rbkt{\tbt{0}{\nu \boldsymbol{C}}{0}{0}\boldsymbol{\Gamma}_T u}}_2}\\
 \leq &~c_0\norm{\boldsymbol{\Gamma}_T\rbkt{\sysf_{\sysp, (\tau+\nu)\sysc} u}}_2+c_0\norm{{\tbt{0}{\nu \boldsymbol{C}}{0}{0}\boldsymbol{\Gamma}_T u}}_2\\
 \leq &~c_0\norm{\boldsymbol{\Gamma}_T\rbkt{\sysf_{\sysp, (\tau+\nu)\sysc}u}}_2+c_0\abs{\nu}\norm{\boldsymbol{C}}\norm{\boldsymbol{\Gamma}_T u}_2,
\end{align*}
where the causality of $\sysc$ and the fact
$\norm{\boldsymbol{\Gamma}_T (\cdot)}_2$ is a nondecreasing function of $T$ are used. The above inequality gives 
$\norm{\boldsymbol{\Gamma}_T u}_2\leq
 \frac{c_0}{ \rbkt{1-c_0\abs{\nu}\norm{\boldsymbol{C}}}}\norm{\boldsymbol{\Gamma}_T\rbkt{\sysf_{\sysp, (\tau+\nu)\sysc} u}}_2$
provided that $\abs{\nu} < {1}/({c_0\norm{\boldsymbol{C}}})\eqqcolon\mu$.

\emph{Step 3}: Show that $\boldsymbol{P}\,\#\,\rbkt{\tau\boldsymbol{C}} |_{e_2=0}$ is stable when $\tau=1$.

When $\tau=0$, $\rbkt{\sysf_{\sysp, \tau\sysc}}^{-1}$ is bounded since $\boldsymbol{P}$ is open-loop stable. It has been shown in {Step 2} that $\rbkt{\sysf_{\sysp, \tau\sysc}}^{-1}$ is bounded for $\tau < \mu$, and then it is bounded for $\tau < 2\mu$ using the iterative process, etc. By induction, $\rbkt{\sysf_{\sysp, \tau\sysc}}^{-1}$ is bounded for all $\tau \in \interval{0}{1}$. Thus, $\gof|_{e_2=0}$ is stable by setting $\tau=1$. \hfill $\square$
\endpf
We call \eqref{eq:spt} the \emph{small angle condition} serving as a counterpart to the elegant small gain one \eqref{eq:small-gain}. Specifically, the former involves a comparison of the loop singular angle $\theta(\sysp)+\theta(\sysc)$ with $\pi$, while the latter a comparison of the loop $\lt$-gain $\norm{\sysp}\norm{\sysc}$ with $1$. Importantly, condition \eqref{eq:spt} provides a new realization of Zames' anticipation \cite[Sec.~6.3]{Zames:66_2} in which a stability condition involving ``loop absolute phase-shift'' has been speculated.

Theorem~\ref{thm:spt} also holds for the other single-loop feedback system $\gof|_{e_1=0}$ defined in the sense of setting the other external signal $e_1=0$ in Fig.~\ref{fig:feedback}. Observe that stronger feedback stability with an ``\emph{infinite gain margin}'' is guaranteed in Theorem~\ref{thm:spt}; namely, if condition (\ref{eq:spt}) holds, then the well-posed $\rbkt{\tau\boldsymbol{P}}\,\#\,\boldsymbol{C}|_{e_2=0}$ is also stable for all $\tau >0$. A one-line proof follows from $\theta(\tau\sysp)=\theta(\sysp)$ for all $\tau>0$. This coincides with the infinite gain margin concept in classical control theory. Moreover, the quantity $\pi-{\theta(\sysp)-\theta(\sysc)}$ gives a new robustness indicator of nonlinear feedback systems. It characterizes a smallest ``\emph{phase margin}'' for $\gof|_{e_2=0}$ over all positive gain uncertainties in the feedback loop, i.e., $\rbkt{\tau\boldsymbol{P}}\,\#\,\rbkt{\nu\boldsymbol{C}}|_{e_2=0}$ for all $\tau, \nu >0$. This may facilitate our understanding of an open question posted in \cite[p.~71]{Sepulchre:97}: What are phase margins of nonlinear feedback systems? 

When $\sysp$ is stable passive in \eqref{eq:stable_passive} and $\sysc$ is very strictly passive in \eqref{eq:VSP}, Theorem~\ref{thm:spt} reduces to a version of the passivity theorem \cite[Sec.~6.6.2]{Vidyasagar:93} by noting $\theta(\sysp)\leq \pi/2$ and $\theta(\sysc)<\pi/2$. To make this clear, we build a link between the singular angle and very strict passivity, as detailed in Section~\ref{sec:05}. In addition, the small angle condition \eqref{eq:spt} is non-quadratic due to \eqref{eq:singular_angle}. To the best of our knowledge, it may not be expressed in terms of integral quadratic constraints (IQCs) and thus Theorem~\ref{thm:spt} may not be recovered from existing IQCs nonlinear feedback stability results, e.g., \cite{Rantzer:97, Khong:21}. We also note a recent notion of the system scaled relative graph (SRG) \cite{Chaffey:21j} which incorporates both gain and angle information in an incremental form. A feedback system analysis is highlighted in \cite{Chaffey:21j} based on SRG properties and assumptions. 
By contrast, the singular angle is non-incremental and may be understood graphically from the maximum angle on the scaled graph \cite{Chaffey:21j}. More importantly, Theorem~\ref{thm:spt} is a novel result that forms the foundation of an angle-based input-output nonlinear theory that well complements the presently dominating gain-based theory.

In Theorem~\ref{thm:spt}, only two open-loop systems are involved in the feedback loop. This does not fully reveal the advantage of the system singular angle in terms of the cascaded property in Proposition~\ref{prop:productPC}. To this end, we apply Theorem~\ref{thm:spt} to commonly-seen cyclic systems.

\subsection{An Application to Cyclic Systems}\label{sec:cyclic}
\begin{figure}[htb]
\centering
\setlength{\unitlength}{0.9mm}
\begin{picture}(80,25)
\thicklines
\put(0,20){\vector(1,0){8}} \put(10,20){\circle{4}}
\put(12,20){\vector(1,0){8}}
\put(20,15){\framebox(10,10){$\sysp_1$}}
\put(30,20){\vector(1,0){5}}
\put(35,15){\framebox(10,10){$\boldsymbol{P}_2$}}
\put(45,20){\vector(1,0){5}}
\put(51,17.5){\makebox(5,5){$\cdots$}}
\put(56,20){\vector(1,0){5}}
\put(61,15){\framebox(10,10){$\sysp_N$}}
\put(71,20){\line(1,0){10}} \put(81,20){\line(0,-1){15}}
\put(81,5){\line(-1,0){71}}
\put(10,5){\vector(0,1){13}}
\put(74,20){\makebox(5,5){$y_N$}}
\put(0,20){\makebox(5,5){$ e$}}
\put(30,20){\makebox(5,5){$ u_2$}}
\put(45,20){\makebox(5,5){$ y_2$}}
\put(55,20){\makebox(5,5){$ u_{N}$}}
\put(13,20){\makebox(5,5){$u_1$}} 
\put(10,10){\makebox(6,10){$-$}}
\end{picture}
\vspace{-2mm}
\caption{A cyclic system.} \label{fig:cyc}
\end{figure}

Consider the feedback system with a special network structure shown in Fig.~\ref{fig:cyc}, which is called a cyclic system, where $\sysp_1, \sysp_2, \ldots, \sysp_N$ are $N$ causal stable subsystems defined on $\ltep$. The cyclic system and its stability condition have been studied in \cite{Sontag:06, Arcak:06} disciplines. A notable stability condition, called the secant condition, is given in \cite{Sontag:06, Arcak:06}. Concretely, for an output strictly passive system $\sysp\colon\ltep \to \ltep$ (see Section~\ref{sec:02}), there exists $\gamma>0$ such that
\be\label{eq:secant_gain}
\gamma\ininf{u}{\syspu} \geq \norm{\syspu}_2^2 \quad \forall u \in \ltp.
\ee
Here $\gamma$ is connected to the output passivity index $\rho$ in a reciprocal relation, i.e., $\gamma=1/\rho$. Then, the smallest $\gamma$ as in (\ref{eq:secant_gain}) is called the secant gain $\gamma_s(\sysp)$ of $\sysp$ in \cite{Sontag:06, Arcak:06}. By introducing this, the author of \cite{Sontag:06} states that the well-posed cyclic system is stable if $\sysp_1, \sysp_2, \ldots, \sysp_N$ are output strictly passive and the following secant gain condition holds:
\be \label{eq:secant_gain_condition}
\textstyle \prod_{i=1}^{N} \gamma_s(\sysp_i) < \rbkt{\sec \frac{\pi}{N}}^N.
\ee
Condition (\ref{eq:secant_gain_condition}), just as its name implies, is more like a gain-flavor result. This is also evidenced by the following two facts in \cite{Sontag:06}: First, for an output strictly passive system $\sysp$, it holds that $\norm{\sysp}\leq \gamma_s(\sysp)$, and thus the secant gain provides an upper bound of the $\lt$-gain. Second, the right-hand side of (\ref{eq:secant_gain_condition}) tends monotonically to $1$ in a decreasing way as $N \to \infty$, which reduces to a special small gain condition.

By virtue of Theorem~\ref{thm:spt}, we here provide an angular stability condition for cyclic systems as a corollary.
\begin{cor}\label{thm:cyclic}
 For causal stable systems $\sysp_1, \sysp_2, \ldots, \sysp_N$, the well-posed cyclic system is stable if
 $\sum_{i=1}^{N}\theta(\sysp_i)<\pi$.
\end{cor}
\pf
 According to the triangle inequality in Lemma~\ref{lem:signal_triangle_ineq} and Proposition~\ref{prop:productPC}, we have $\theta(u_1, y_N)>-\pi$. The theorem then can be proved using the same arguments as in the proof of Theorem~\ref{thm:spt}. \hfill $\square$
\endpf
 
\section{The Singular Angle of a MIMO LTI System and an LTI Small Angle Theorem}\label{sec:07}
An LTI system can be viewed as a convolution operator in the time domain or as a transfer function matrix in the frequency domain. In general, there are two routes to defining the singular angle of a MIMO LTI system. One is to directly inherit the time-domain singular angle definition in Section~\ref{sec:03}, with the system being LTI. The other is to define the frequency-domain singular angle of transfer function matrices by means of the matrix singular angle. But are these two approaches equivalent? The answer is generally negative, as elaborated in the following simplest case of SISO LTI systems. 

Given a SISO LTI system $\sysp$ with $P(s)\in \rhinf^{1\times 1}$, we define the \textit{$\hinf$ singular angle} of $P(s)$ as:
\be\label{eq:siso_LTI_sphase}
{\theta_\infty}(P) \coloneqq \sup_{\substack{ \omega\in \interval{0}{\infty}, {P(\jw)\neq 0}}} \arccos \frac{\rep \rbkt{P(\jw)}}{\abs{P(\jw)}}.
\ee
To avoid any ambiguity, we use the subscript ``$\infty$'' to indicate the $\hinf$ singular angle $\theta_\infty(P)$ in the frequency domain, which distinguishes it from the singular angle $\theta(\sysp)$ defined by (\ref{eq:nonlinear_angle}) in the time domain. Notice that the term in (\ref{eq:siso_LTI_sphase}) can be interpreted via $\arccos {\frac{\rep \rbkt{P(\jw)}}{\abs{P(\jw)}}}= \abs{\angle P(\jw)}$ when $P(\jw)\neq 0$, where $\angle P(\jw)$ is the classical phase response of $P(s)$. Roughly speaking, the $\hinf$ singular angle of $P(s)$ returns the ``largest phase'' of $P(s)$. The appellation ``$\hinf$ singular angle'' is motivated by the well-known $\hinf$-norm which returns the ``largest gain'' of $P(s)$. The following proposition connects $\theta_\infty(P)$ and $\theta(\sysp)$, whose proof is provided in Appendix.  
\begin{prop}\label{prop:SISO}
For a SISO LTI system $\sysp$ with $P(s)\in \rhinf^{1\times 1}$, the following two statements are true:
\begin{enumerate}
 \renewcommand{\theenumi}{\textup{(\roman{enumi})}}\renewcommand{\labelenumi}{\theenumi}
 \item \label{item: SISO_angle} ${\theta_\infty}(P)\leq \theta(\sysp)$.
 \item \label{item: SISO_angle_2} If $\theta_\infty(P) \in \interval[open left]{\pi/2}{\pi}$, then $\theta_\infty(P)=\theta(\sysp)$.
\end{enumerate}
\end{prop}

It is noteworthy that for a certain SISO LTI system, when $\theta_\infty(P) <{\pi/2}$, it actually holds that $\theta_\infty(P)<\theta(\sysp)$. A concrete example is provided in Appendix to support this claim. From the above discussion on SISO LTI systems, it is meaningful to develop a singular angle notion for LTI systems via a frequency-domain approach, since frequency-wise analysis is a key feature in LTI systems theory. In the remainder of this section, we first lay the mathematical foundations by investigating the matrix singular angle and then address MIMO LTI systems on the shoulders of the matrix foundation. 

\subsection{The Matrix Singular Angle}
The angle definition \eqref{eq:singular_angle} between elements in $\ltp$ can be naturally extended to a Hilbert space $\mathcal{H}$. Specifying $\mathcal{H}=\mathbb{C}^n$ endowed with a real-valued inner product will bring the following angle between complex vectors. For vectors $x, y\in \mathbb{C}^n$, the angle $\theta(x, y)\in \interval{0}{\pi}$ between $x$ and $y$ is defined via
\bex
\theta(x, y)\coloneqq \displaystyle \arccos \frac{\rep \rbkt{x^* y}}{\abs{x}\abs{y}},
\quad \text{if}~x,y \neq 0,
 \eex
and $\theta(x, y)\coloneqq 0$, if $x=0$ or $y=0$. For a nonzero matrix $A\in \mathbb{C}^{n\times n}$, the singular angle $\theta(A)\in \interval{0}{\pi}$, introduced in \cite[Sec.~23]{Wielandt:67} and \cite[Ch.~3]{Gustafson:97}, is defined by
\be\label{eq:matrix_angle}
\theta(A)\coloneqq \displaystyle \sup_{ {0\neq x\in \mathbb{C}^n, {Ax}\neq 0}} \theta(x, Ax).
\ee
The following lemma has been proved in {\cite[Sec.~23.5]{Wielandt:67}} and can also be deduced from Proposition~\ref{prop:productPC} by changing $\ltp$ to $\mathbb{C}^n$.
\begin{lem}\label{lem:matrix_triangle_ineq}
 For nonzero matrices $A, B\in \mathbb{C}^{n\times n}$, it holds that
 $\theta(BA)\leq \theta(A)+\theta(B)$.
\end{lem}
 
Let $\lambda_i(A)\in \mathbb{C}$ denote the $i$-th eigenvalue of $A$, where $i=1,2, \ldots, n$. The next lemma is a modification of the result in {\cite[Sec.~23.7]{Wielandt:67}}.
\begin{lem}\label{lem:spectrum_sangle}
 For a nonzero matrix $A\in \mathbb{C}^{n\times n}$, if $\theta(A)<\pi$, then $\abs{\angle\lambda_i(A)}<\pi$, where $\lambda_i(A)\neq 0$.
\end{lem}
 
For matrices $A, B\in \mathbb{C}^{n\times n}$, the singularity of $I+BA$ is an essential issue when we study MIMO LTI feedback systems. Combining Lemmas~\ref{lem:matrix_triangle_ineq}~and~\ref{lem:spectrum_sangle} yields an angular condition for the purpose of determining the singularity of $I+BA$, as detailed in the following theorem.
\begin{thm}
For nonzero $A, B\in \mathbb{C}^{n\times n}$, it holds that $\det(I+BA)\neq0$ if
 ${\theta}(A)+{\theta}({B})<\pi$.
\end{thm}
\pf
Using Lemma~\ref{lem:matrix_triangle_ineq}, we have that $\theta(BA)\leq \theta(A)+\theta(B)<\pi$.
 According to Lemma~\ref{lem:spectrum_sangle}, we obtain that $\angle \lambda(BA)\in \interval[open]{-\pi}{\pi}$, which gives $\det(I+BA)\neq0.$ \hfill $\square$
\endpf

Equipped with the matrix singular angle, we are ready to cope with feedback stability of MIMO LTI systems.

\subsection{The Singular Angles of MIMO LTI Systems}
Consider a MIMO LTI system $\sysp$ with ${P}(s)\in \rhinf^{n\times n}$. Recall that, $\overline{\sigma}(P(\jw))$, the largest singular value of $P(\jw)$, is a function of the frequency $\omega\in\interval{0}{\infty}$. Then $\overline{\sigma}(P(\jw))$ is often called the frequency-wise gain of $P(s)$ and $\hinf$-norm of $P(s)$ is given by
$\norm{P}_\infty\coloneqq \sup_{\omega\in \interval{0}{\infty}} \overline{\sigma}(P(\jw))$.

The MIMO LTI system singular angle can be defined in a similar way. We define the \textit{frequency-wise singular angle} ${\theta}(P(\jw)) \in \interval{0}{\pi}$ for each frequency $\omega\in\interval{0}{\infty}$ based on the matrix singular angle of $P(\jw)$ in definition~(\ref{eq:matrix_angle}), i.e.,
\bex
{\theta}(P(\jw)) \coloneqq \sup_{\substack{0\neq x\in \mathbb{C}^n, {P(\jw)x}\neq 0}} \theta(x, P(\jw)x),~\text{if}~ P(\jw)\neq 0, 
\eex
and ${\theta}(P(\jw))\coloneqq 0$, otherwise. Additionally, the \textit{$\hinf$ singular angle} ${\theta_\infty}(P) \in \interval{0}{\pi}$ is then defined to be 
\be\label{eq:fr_LTI_sphase}
{\theta_\infty}(P) \coloneqq \sup_{\omega\in \interval{0}{\infty}}\theta(P(\jw)).
\ee
Clearly, \eqref{eq:fr_LTI_sphase} generalizes the SISO case presented in \eqref{eq:siso_LTI_sphase}.

We next investigate the feedback stability of MIMO LTI systems. For LTI systems $\sysp$ and $\sysc$ with $P(s), C(s) \in \rhinf^{n\times n}$, the well-posedness and feedback stability definitions in Section~\ref{sec:04} need further clarification and simplification. Specifically, the well-posedness of LTI feedback system $\gof$ is equivalent to $\rbkt{I+C(s)P(s)}^{-1}$ existing and being proper \cite[Lem.~5.1]{Zhou:96}. Since $P(s), C(s) \in \rhinf^{n\times n}$, then according to \cite[Cor.~5.6]{Zhou:96}, $\gof$ is stable if and only if $\rbkt{I+C(s)P(s)}^{-1} \in \rhinf^{n\times n}$.

Recall a frequency-wise version of the small gain theorem \cite{Zhou:96} for guaranteeing feedback stability of LTI $\gof$:
$\overline{\sigma}(P(\jw)) \overline{\sigma}(C(\jw)) <1$ for all $\omega\in\interval{0}{\infty}$. We now provide a counterpart below. 

\begin{thm}[Frequency-wise small angle theorem]\label{thm:frequency-wise_sptm}
Let $\sysp$ and $\sysc$ be LTI systems with $P(s), C(s)\in \rhinf^{n\times n}$. The well-posed $\gof$ is stable if 
 \bex
 {\theta}(P(\jw)) +{\theta}(C(\jw)) <\pi \quad \forall \omega \in \interval{0}{\infty}.
 \eex
\end{thm}
\pf
In light of \cite[Cor.~5.6]{Zhou:96}, when $P(s)$ and $C(s)$ are stable, it suffices to show that $\det \sbkt{I+C(s)P(s)}\neq 0$ for all $s\in \ccp \cup \{\infty\}$. We adopt a homotopy method by letting $\tau$ be an arbitrary number in $\interval{0}{1}$. Note that $\theta(C(\jw))=\theta(\tau C(\jw))$ when $\tau\in\interval[open left]{0}{1}$ and $\theta(C(\jw))\geq \theta(\tau C(\jw))=0$ when $\tau=0$. By Lemma~\ref{lem:matrix_triangle_ineq} and hypothesis, for all $\omega\in \interval{0}{\infty}$ and all $\tau\in \interval{0}{1}$, we have
 \begin{align*}
 {\theta}(\tau C(\jw)P(\jw))&\leq {\theta}(P(\jw))+{\theta}(\tau C(\jw))\\
 &\leq {\theta}(P(\jw))+{\theta}(C(\jw))<\pi.
 \end{align*}
 It follows from Lemma~\ref{lem:spectrum_sangle} that, for all $\tau\in \interval{0}{1}$, we have
 \bex
 -\pi<\angle\lambda_i(\tau C(\jw)P(\jw))<\pi\quad \forall \omega \in \interval{-\infty}{\infty}
 \eex
for $i=1,2, \ldots, n$. This gives that, for all $\tau\in \interval{0}{1}$,
 \be\label{eq:LTI_detneq0jR}
 \det (I+\tau C(\jw)P(\jw)) \neq 0\quad \forall \omega \in \interval{-\infty}{\infty}.
 \ee
We then extend the result on the imaginary axis to the closed right half-plane. When $\tau=0$, it holds that $\det (I+\tau C(s)P(s)) \neq 0$ for all $s\in \ccp \cup \{\infty\}$.
According to the continuity of the closed-loop system poles and by (\ref{eq:LTI_detneq0jR}), we obtain that 
 $\det (I+\tau C(s)P(s)) \neq 0$ for all $s\in \ccp \cup \{\infty\}$ and $\tau \in \interval{0}{1}$. The proof is completed by setting $\tau=1$. \hfill $\square$
\endpf
For MIMO LTI systems, one may verify that \ref{item: SISO_angle} of Proposition~\ref{prop:SISO} still holds since the proof can be adapted to a tailored vector signal $u\in \mathcal{L}_2^n$ such that $u=xu_0$, where $x\in \cn$ and $u_0\in \mathcal{L}_2^1$ is constructed in \eqref{eq: unit_signal}. It follows from definition~\eqref{eq:fr_LTI_sphase} that the following relation holds:
\be\label{eq:three_angles}
\theta(P(\jw)) \leq \theta_\infty(P) \leq \theta(\sysp)\quad \forall \omega\in \interval{0}{\infty}.
\ee
Consequently, the use of $\theta(P(\jw))$, in contrast to that of $\theta(\sysp)$, in formulating the LTI small angle theorem can reduce conservatism. The next corollary involves the $\mathcal{H}_\infty$ singular angle, which follows directly from Theorem~\ref{thm:frequency-wise_sptm} by noting~\eqref{eq:three_angles}.
 \begin{cor}\label{coro:hinf_sat}
 Let $\sysp$ and $\sysc$ be LTI systems with $P(s), C(s)\in \rhinf^{n\times n}$. The well-posed $\gof$ is stable if
 ${\theta_\infty}(P) +{\theta_\infty}(C) <\pi$.
 \end{cor}
The condition above complements the small gain one: $\norm{P}_\infty\norm{C}_\infty<1$.

\subsection{A Geometric Interpretation}
We facilitate the geometric understanding of the small angle theorem (Theorem~\ref{thm:spt}) using a SISO LTI feedback system $\gof$ against ``conic uncertainty''. Suppose that the Nyquist plot of an uncertain system $\sysp$ lives in a cone such that $\abs{\angle P(\jw)}<\theta(\sysp)<\pi/2$ and that of an uncertain system $\sysc$ lives in a cone such that $ \theta(\sysc)>\abs{\angle C(\jw)}>\pi/2$. In this case, the ``conic uncertainty'' may exist in both $\sysp$ and $\sysc$, as shown in Fig.~\ref{fig:sam_interpretation}. If the small angle condition (\ref{eq:spt}) is satisfied, thereby existing a tradeoff between $\theta(\sysp)$ and $\theta(\sysc)$, then it follows from Proposition~\ref{prop:SISO} that, for all $\omega\in\interval{-\infty}{\infty}$, we have
\begin{align*}
\abs{\angle P(\jw)C(\jw)}&\leq \sup_{\omega\in\interval{0}{\infty}} \abs{\angle P(\jw)} + \sup_{\omega\in\interval{0}{\infty}} \abs{\angle C(\jw)}\\
&< \theta(\sysp)+\theta(\sysc)< \pi.
\end{align*}
From the above inequality, there is no intersection between the Nyquist plot of $\sysp\sysc$ and the negative real axis. This feedback system $\gof$ is obviously stable which falls under the classical Nyquist criterion. Now, we have generalized the above stability result to MIMO LTI systems (Theorem~\ref{thm:frequency-wise_sptm} and Corollary~\ref{coro:hinf_sat}) and even to nonlinear systems (Theorem~\ref{thm:spt}).

\begin{figure}[htb]
 \centering
 \setlength{\unitlength}{0.8mm}
 \begin{picture}(100,40)
 \put(5,20){\vector(1,0){40}}
 \put(30,5){\vector(0,1){35}}
 {\thicklines
 \put(33,18){\makebox(0,0){$0$}}
 \multiput(30,20)(-4,3){6}{\line(-4,3){3}}
 \multiput(30,20)(-4,-3){6}{\line(-4,-3){3}}
 \thicklines
 {\color{red}\put(29, 20){\arc[0,138]{6}}
 \color{red}\put(29, 20){\arc[-138,0]{6}}}
 \put(43,17){\makebox(0,0){$\rep$}}
 \put(14,38){\makebox(0,0){$\theta(\sysc)$}}
 \put(26,38){\makebox(0,0){$\imp$}}}
 \put(22,0){\makebox(0,0){${\theta(\sysc)>}\abs{\angle C(\jw)}>\pi/2$}}
 \put(55,20){\vector(1,0){40}}
 \put(70,5){\vector(0,1){35}}
 \thicklines
 \put(68,18){\makebox(0,0){$0$}}
 \multiput(70,20)(4,3){6}{\line(4,3){3}}
 \multiput(70,20)(4,-3){6}{\line(4,-3){3}}
 \thicklines
 {\color{red}\put(69, 20){\arc[0,34]{10}}
 \color{red}\put(69, 20){\arc[-34,0]{10}}
 }
 \put(93,17){\makebox(0,0){$\rep$}}
 \put(93,30){\makebox(0,0){$\theta(\sysp)$}}
 \put(66,38){\makebox(0,0){$\imp$}}
 \put(77,0){\makebox(0,0){$\abs{\angle P(\jw)}{ < \theta(\sysp)}<\pi/2$}}
 \end{picture} \caption{An illustration of a SISO LTI $\gof$ against ``conic uncertainty''.} \label{fig:sam_interpretation}
 \end{figure}

\section{Relation to the Passivity}\label{sec:05}
In this section, we establish a connection between the system singular angle and the well-known notion of passivity. As a prologue, we first estimate the singular angle of a very strictly passive system. Second, we demonstrate that a system with a singular angle greater than $\pi/2$ can be equivalently represented by an input-feedforward-output-feedback passive system with a group of constraints. Third, we apply the small angle theorem to a Lur'e system by estimating the singular angle of a sector bounded static nonlinearity, and provide an angular interpretation of the celebrated circle criterion with an ``infinite gain margin''.

\subsection{Relation to the Input/Output Passivity Indices}
 
The following proposition shows that the worst case of the singular angle of a very strictly passive system can be estimated from its given passivity indices as a bound. 
\begin{prop}\label{prop:angle_VSP}
 For a very strictly passive system $\sysp$ in \eqref{eq:VSP} associated with given indices $\nu, \rho>0$, it holds that
\be\label{eq:VSP_phase}
\theta(\sysp)\leq \arccos2\sqrt{\nu\rho}<\pi/2.
\ee
\end{prop}

\pf
For all $u \in \ltp\setminus\{0\}$ with $\sysp u\neq 0$, rearranging inequality (\ref{eq:VSP}) yields that
 $\frac{\ininf{u}{\sysp u}}{\norm{u}_2\norm{\sysp u}_2} \geq \nu \frac{\norm{u}_2}{\norm{\sysp u}_2}+ \rho \frac{\norm{\sysp u}_2}{\norm{u}_2}$.
The right-hand side of the above inequality is bounded from below by a constant according to the geometric and arithmetic means inequality, namely,
\bex
\nu \frac{\norm{u}_2}{\norm{\sysp u}_2}+ \rho \frac{\norm{\sysp u}_2}{\norm{u}_2}\geq 2\sqrt{\nu\rho\frac{\norm{u}_2}{\norm{\sysp u}_2}\frac{\norm{\sysp u}_2}{\norm{u}_2} }=2\sqrt{\nu\rho}.
\eex
This gives 
$\cos\theta(\sysp)=\inf_{\substack{ 0\neq u \in \ltp,\\{\sysp u\neq 0}}}\frac{\ininf{u}{\boldsymbol{P}u}}{\norm{u}_2\norm{\sysp u}_2}\geq 2\sqrt{\nu\rho}$.
Notice that there is an implicit constraint for the very strict passivity that $\nu$ and $\rho$ always satisfy $\nu\rho\leq {1}/{4}$. We then conclude that $\theta(\sysp)\leq \arccos 2\sqrt{\nu\rho}<\pi/2$. \hfill $\square$
\endpf 

It is noteworthy that the singular angle depends solely on the \emph{product} $\nu\rho$ rather than the two individual indices. This observation is intuitive, as the indices provide two parameters to characterize the system, whereas the singular angle -- like the $\lt$-gain -- represents only a single parameter. The next proposition reveals that a system whose singular angle is greater than $\pi/2$ is an input-feedforward-output-feedback passive system satisfying a group of constraints.
\begin{prop}\label{prop:IFOFP}
Let $\sysp$ be causal and stable and $\alpha\in \interval[open left]{\pi/2}{\pi}$. The following two statements are equivalent:
\begin{enumerate}
 \renewcommand{\theenumi}{\textup{(\roman{enumi})}}\renewcommand{\labelenumi}{\theenumi}
 \item \label{item:prop_IFOFP_1} $\theta(\sysp)\in \interval[open left]{\pi/2}{\alpha}$.
 \item \label{item:prop_IFOFP_2} For all $\nu,\rho<0$ satisfying $\nu\rho=\frac{\rbkt{\cos\alpha}^2}{4}$, it holds that
 \be\label{eq:lem_IFOFP_1}
 \ininf{u}{\syspu}\geq \nu\norm{u}^2_2+ \rho\norm{\syspu}_2^2\quad \forall u\in\ltp.
 \ee
 \end{enumerate}
\end{prop}
\pf
 \ref{item:prop_IFOFP_1} $\to$ \ref{item:prop_IFOFP_2}: By (\ref{eq:nonlinear_angle}), for all $u\in \ltp$, we have
\begin{align*}
 {\ininf{u}{\sysp u}}& \geq \cos\theta(\sysp)\norm{u}_2\norm{\syspu}_2\geq \cos\alpha \norm{u}_2\norm{\syspu}_2\\
 &\geq \frac{\cos\alpha}{2} \rbkt{c\norm{u}_2^2+\frac{1}{c}\norm{\sysp u}^2_2}\quad \forall c>0,
\end{align*}
 where the second and last inequalities use the assumption $0>\cos\theta(\sysp)\geq \cos\alpha$ and the geometric and quadratic
means inequality, respectively. Therefore, the indices $\nu$ and $\rho$ are parameterized by $c$ in the following way:
 $\nu=\frac{c\cos\alpha}{2}<0$ and $\rho=\frac{\cos\alpha}{2c}<0$ for all $c>0$.

 \ref{item:prop_IFOFP_2} $\to$ \ref{item:prop_IFOFP_1}: All we need to show is that, for all $u\in\ltp\setminus\{0\}$ with $\sysp u\neq 0$, we have ${\ininf{u}{\boldsymbol{P}u}} \geq \cos\alpha {\norm{u}_2\norm{\sysp u}_2}$. To this end, let $u$ be arbitrary in $\ltp\setminus\{0\}$ with $\sysp u\neq 0$. Then, we choose that
\bex
\nu=\frac{\cos\alpha\norm{\sysp u}_2}{2\norm{u}_2}<0\quad\text{and}\quad \rho=\frac{\cos\alpha\norm{u}_2}{2\norm{\sysp u}_2}<0.
\eex
Clearly, $\nu\rho={\rbkt{\cos\alpha}^2}/{4}$. Then, assertion~\ref{item:prop_IFOFP_2} tells that
\begin{align*}
 {\ininf{u}{\sysp u}} &\geq \frac{\cos\alpha\norm{\sysp u}_2}{2\norm{u}_2}\norm{u}^2_2+ \frac{\cos\alpha\norm{u}_2}{2\norm{\sysp u}_2}\norm{\syspu}_2^2 \\
 &=\cos\alpha {\norm{u}_2\norm{\sysp u}_2}. \hspace{3.7cm}\square
\end{align*} 
\endpf
In Proposition~\ref{prop:IFOFP}, similarly the indices $\nu$ and $\rho$ should meet $\nu\rho=\frac{\rbkt{\cos\alpha}^2}{4}$; i.e., their product is a constant. This gives an input-feedforward-output-feedback passive system satisfying a set of constraints parameterized by $\alpha$. Roughly speaking, a system's singular angle greater than $\pi/2$ can be understood via merging a group of the two constrained indices together. See Fig.~\ref{fig:passive_angle} for a graphical illustration.
 
\begin{figure}[htb]
 \centering
 \setlength{\unitlength}{0.9mm}
 \begin{picture}(60,40)
 \put(0,20){\vector(1,0){55}}
 \put(40,5){\vector(0,1){35}}
 \thicklines
 \put(43,18){\makebox(0,0){$0$}}
 {\thinlines\color{blue}\put(22.5,20){\circle{20.5}}
 \put(31,20){\circle{10.5}}
 \put(16,20){\circle{28.5}}
 }
 \put(45,22){\makebox(0,0){{\color{red}$\alpha$}}}
 \put(22.5,16.5){\makebox(0,0){$\frac{1}{2\rho}$}}
 \put(20,25){\makebox(0,0){$\frac{\sqrt{1-4\nu\rho}}{\abs{2\rho}}$}}
 \multiput(40,20)(-4,3){7}{\line(-4,3){3}}
 \multiput(40,20)(-4,-3){7}{\line(-4,-3){3}}
 \thicklines
 \put(22.5,20){\vector(3,4){6.3}}
 {\color{red}\put(39, 20){\arc[0,128]{3}}}
 \put(53,17){\makebox(0,0){$\rep$}}
 \put(36,38){\makebox(0,0){$\imp$}}
 \end{picture} \caption{A graphical illustration of Proposition~\ref{prop:IFOFP} when $\sysp$ is SISO LTI with $P(s)\in \rhinf^{1\times 1}$. It is easy to verify that the constraint (\ref{eq:lem_IFOFP_1}) can be equivalently depicted by the set outside blue disks in $\mathbb{C}$ parameterized by $\nu$ and $\rho$ with the center $\frac{1}{2\rho}$ and radius $\frac{\sqrt{1-4\nu\rho}}{\abs{2\rho}}$, where the product $\nu\rho=\frac{\rbkt{\cos\alpha}^2}{4}$ is fixed. This set forms a cone (two dashed rays) which opens to the right. Notably, this cone is exactly characterized by the singular angle information $\alpha$.} \label{fig:passive_angle}
\end{figure}

\subsection{An Angular Interpretation of the Circle Criterion}
We aim at showing an angular interpretation of the famous circle criterion under an ``infinite gain margin'' constraint based on the small angle theorem. 

Consider a scalar static nonlinear system $\sysn: \ltep \rightarrow \ltep$ defined by
$(\sysn u)(t)=h(u(t))$, where $h: \mathbb{R}\rightarrow\mathbb{R}$ satisfies
\be\label{eq:sector}
\rbkt{h(x)-a x}\rbkt{h(x) -b x } \leq 0,\quad \forall x \in \mathbb{R}
\ee with $b > a > 0$. Such an $\sysn$ is called a sector bounded static nonlinearity and belongs to the {nonlinearity sector} from $a$ to $b$. It is known that $\sysn$ is very strictly passive. To observe this, according to \eqref{eq:sector}, for all $u \in \ltp\setminus\{0\}$ and for all $t\geq 0$, we have $\ininf{u}{\boldsymbol{N}u}\geq \frac{ab}{a+b}\norm{u}^2_2 +\frac{1}{a+b} \norm{\boldsymbol{N}u}^2_2$.
In light of \eqref{eq:VSP}, the passivity indices are $\nu=\frac{ab}{a+b}$ and $\rho=\frac{1}{a+b}$. Therefore, we can estimate $\theta(\sysn)$ from the sector parameters $a$ and $b$, as detailed in the following corollary whose proof follows directly from Proposition~\ref{prop:angle_VSP}.
\begin{cor}\label{prop:nonlinearity}
 For a static system $\sysn$ satisfying the sector condition (\ref{eq:sector}), we have
 $\theta(\sysn)\leq \arccos \frac{2\sqrt{ab}}{a+b}<\frac{\pi}{2}$.
\end{cor}

Consider the Lur'e system $\sysp\,\#\,{\sysn}$ consisting of a SISO LTI system $\sysp$ with $P(s)\in \rhinf^{1\times 1}$ and a scalar static system $\boldsymbol{N}$ satisfying \eqref{eq:sector}. The aim is to derive a stability condition on $P(s)$ against all static nonlinearities contained in a sector from $a$ to $b$, a.k.a. absolute stability \cite{Carrasco:16}. For such a feedback system $\sysp\,\#\,{\sysn}$, the celebrated circle criterion \cite[Sec.~6.6.1]{Vidyasagar:93} has stood out as being endowed with a nice geometric interpretation. Specifically, $P(s)$ is required to meet
\bex
\inf_{{\omega\in\interval{-\infty}{\infty}, z\in D\rbkt{a, b}}} \abs{P(\jw)-z}>0,
\eex
where $D \rbkt{a, b} \coloneqq \bbkt{z\in \mathbb{C} | ~\abs{z + \frac{a+b}{2ab}} \leq \frac{b-a}{2ab}}$ denotes a disk.
In other words, the Nyquist plot of $P(s)$ is bounded away from the disk $D \rbkt{a, b}$. See Fig.~\ref{fig:circle} for an illustration. 

\begin{figure}[htb]
\centering
\setlength{\unitlength}{0.8mm}
\begin{picture}(60,40)
\put(0,20){\vector(1,0){55}}
\put(40,5){\vector(0,1){35}}
\thicklines
\put(43,18){\makebox(0,0){$0$}}
\put(0,5){\makebox(5,5){$D\rbkt{a, b}$}}
{\color{blue}\put(22.5,20){\circle{20.5}}}
\put(8,16.5){\makebox(0,0){$-\frac{1}{a}$}}
\put(36,13.5){\makebox(0,0){$-\frac{1}{b}$}}
\put(22.5,16.5){\makebox(0,0){$-\frac{b+a}{2ab}$}}
\put(20,25){\makebox(0,0){$\frac{b-a}{2ab}$}}
\put(56,26){\makebox(0,0){$\pi-\arccos \frac{2\sqrt{ab}}{a+b}$}}
\put(56,10){\makebox(0,0){$\arccos \frac{2\sqrt{ab}}{a+b}-\pi$}}
\multiput(40,20)(-4,3){6}{\line(-4,3){3}}
\multiput(40,20)(-4,-3){6}{\line(-4,-3){3}}
\thicklines
\put(22.5,20){\vector(3,4){6.3}}
{\color{red}\put(39, 20){\arc[0,122]{3}}
\put(38.5, 20){\arc[-120,0]{3}}}
\put(53,17){\makebox(0,0){$\rep$}}
\put(36,38){\makebox(0,0){$\imp$}}
\end{picture}\caption{The interpretations of the circle criterion with the disk $D(a, b)$ (blue) and the small angle theorem with two rays (dashed) for a SISO Lur'e system when $b>a>0$.}\label{fig:circle}
\end{figure}

We now figure out an angular interpretation of the circle criterion. It is known that both magnitude and angle information of $P(s)$ is utilized in the circle criterion. {However, only angle information is considered in Theorem~\ref{thm:spt}. To obtain a graphical understanding of the circle criterion based on Theorem~\ref{thm:spt},} we rule out the magnitude part by imposing a stronger stability requirement on ${\sysp}\,\#\,{\sysn}$; namely, $\rbkt{\tau {\sysp}}\,\#\,{\sysn}$ should be stable for all $\tau>0$. Roughly speaking, an ``infinite gain margin'' is required. In this case, the disk $D
\rbkt{a, b}$ will be enlarged and become a convex cone formed by infinitely many disks. This cone, shown in Fig.~\ref{fig:circle}, is equivalently characterized by an angular condition in the following corollary rooted in Theorem~\ref{thm:spt}. The proof is provided in Appendix. 

\begin{cor}\label{coro:lure}
 The well-posed Lur'e system $\rbkt{\tau {\sysp}}\,\#\,{\sysn}$ is stable for all $\tau >0$ if
$\angle P(\jw) \in \interval[open]{\arccos \frac{2\sqrt{ab}}{a+b}-\pi}{\pi-\arccos \frac{2\sqrt{ab}}{a+b}}
$~for all $\omega\in\interval{-\infty}{\infty}$ and $P(\jw)\neq 0$.
\end{cor}

\section{Singular Angles of Interconnected Systems}\label{sec:06}

A complex nonlinear network is often composed of a large number of subsystems. For such a network, it is desirable to have scalable analysis; i.e., the property of a network can be deduced from the properties of its subsystems. We have shown in Proposition~\ref{prop:productPC} the advantage of the singular angle in studying cascaded interconnected systems. In this section, we further study the singular angles of feedback and parallel interconnected systems from those of the subsystems, respectively.

For a well-posed feedback system $\gof|_{e_2=0}$ in Fig.~\ref{fig:feedback}, denote the closed-loop map from $e_1$ to $y_1$ by $\boldsymbol{G}\coloneqq e_1 \mapsto y_1:\ltep \rightarrow \ltep$. When the systems $\sysp$ and $\sysc$ have singular angles of distinct upper bounds, the following proposition indicates that the worst case of $\theta(\boldsymbol{G})$ can be estimated from $\theta(\sysp)$ and $\theta(\sysc)$ as an upper bound.
 
\begin{prop}\label{prop:large_angle}
For a stable feedback system $\gof|_{e_2=0}$, assume $\theta(\sysp)+\theta(\sysc)\leq \pi$. Then 
$\theta(\boldsymbol{G})\leq \max\bbkt{\theta(\sysp),\theta(\sysc)}$.
\end{prop}
\pf
By hypothesis, the closed-loop map $\boldsymbol{G}=e_1\mapsto y_1$ is stable and the following two inequalities hold for the signals in Fig.~\ref{fig:feedback}: $\ininf{u_1}{y_1} \geq \cos\theta(\sysp) \norm{u_1}_2 \norm{y_1}_2$ and $\ininf{u_2}{y_2} \geq \cos\theta(\sysc) \norm{u_2}_2 \norm{y_2}_2$
for all $u_1, u_2\in \ltp\setminus\{0\}$ and $y_1, y_2\neq 0$. Adding the inequalities together and using $u_1=e_1-y_2, u_2=y_1$ yield
\be\label{eq:thm2_1}
 \frac{\ininf{e_1}{y_1}}{\norm{y_1}_2} \geq \cos\theta(\sysp) \norm{e_1-y_2}_2+\cos\theta(\sysc)\norm{y_2}_2.
\ee
The condition $\theta(\sysp)+\theta(\sysc)
 \leq \pi$ implies that $\cos\theta(\sysp)+\cos\theta(\sysc)\geq 0$. We prove three possible cases separately.

Case (a): $\theta(\sysp)>\pi/2$. This gives that $\cos\theta(\sysp)<0$ and $ \cos\theta(\sysc)>0$. Therefore, we obtain that 
\begin{align}
\hspace{-0.7em} & \cos\theta(\sysp) \norm{e_1-y_2}_2+\cos\theta(\sysc)\norm{y_2}_2 \notag\\
\hspace{-0.7em} =& \cos\theta(\sysp)\rbkt{\norm{e_1-y_2}_2-\norm{y_2}_2}\notag \\
\hspace{-0.7em} & +\sbkt{\cos\theta(\sysc)+\cos\theta(\sysp)}\norm{y_2}_2\notag \\
 \hspace{-0.7em} \geq & \cos\theta(\sysp)\rbkt{\norm{e_1-y_2}_2-\norm{y_2}_2} \geq \cos\theta(\sysp)\norm{e_1}_2, \label{eq:thm2_2}
\end{align}
where the first inequality comes from discarding the positive term and the last inequality uses the triangle inequality $\norm{e_1-y_2}_2\leq \norm{e_1}_2+\norm{y_2}_2$. Combining (\ref{eq:thm2_1}) and (\ref{eq:thm2_2}) gives that
$\cos\theta(\boldsymbol{G})=\inf_{\substack{ 0\neq e_1 \in \ltp,\\{y_1\neq 0}}}\frac{\ininf{e_1}{y_1}}{\norm{e_1}_2\norm{y_1}_2} \geq \cos\theta(\sysp)$,
which implies $\theta(\boldsymbol{G})\leq \theta(\sysp)$.

Case (b): $\theta(\sysc)>\pi/2$. By the same reasoning, we have
\begin{align*}
 &\cos\theta(\sysp) \norm{e_1-y_2}_2+\cos\theta(\sysc)\norm{y_2}_2\\
 =&\sbkt{\cos\theta(\sysp)+\cos\theta(\sysc)}\norm{e_1-y_2}_2\\
 &+\cos\theta(\sysc)\rbkt{\norm{y_2}_2-\norm{e_1-y_2}_2} \\
 \geq& \cos\theta(\sysc)\rbkt{\norm{y_2}_2-\norm{e_1-y_2}_2} \geq \cos\theta(\sysc)\norm{e_1}_2,
\end{align*}
where the last inequality uses the fact that $-\norm{e_1-y_2}_2\leq \norm{e_1}_2-\norm{y_2}_2$. This gives $\theta(\boldsymbol{G})\leq \theta(\sysc)$.

Case (c): $\theta(\sysc), \theta(\sysp)\leq \pi/2$. Without loss of generality (WLOG), let $\theta(\sysp)\geq \theta(\sysc)$. This gives that $0\leq \cos\theta(\sysp)\leq \cos\theta(\sysc)$. Then we have
$\cos\theta(\sysp) \norm{e_1-y_2}_2+\cos\theta(\sysc)\norm{y_2}_2 
\geq \cos\theta(\sysp) \rbkt{\norm{e_1-y_2}_2+\norm{y_2}_2}$. It follows from \eqref{eq:thm2_1} that 
\begin{align*}
 {\ininf{e_1}{y_1}}/{\norm{y_1}_2} &\geq \cos\theta(\sysp) \rbkt{\norm{e_1-y_2}_2+\norm{y_2}_2} \\
 &\geq \cos\theta(\sysp) \norm{e_1}_2,
\end{align*}
where the last inequality uses the triangle inequality. We then have 
$\cos\theta(\boldsymbol{G})=\inf_{\substack{ 0\neq e_1 \in \ltp,\\{y_1\neq 0}}}\frac{\ininf{e_1}{y_1}}{\norm{e_1}_2\norm{y_1}_2} \geq \cos\theta(\sysp)$
and $\theta(\boldsymbol{G})\leq \theta(\sysp)$. Therefore, for all three cases, we conclude that $
 \theta(\boldsymbol{G})\leq \max\bbkt{\theta(\sysp),\theta(\sysc)}$. \hfill $\square$
\endpf

Given $\alpha\in \interval[open right]{0}{\pi/2}$, let $\mathcal{P}(\alpha)$ be the set of angle-bounded systems defined by $\mathcal{P}(\alpha)\coloneqq \bbkt{\sysp:\ltep \rightarrow \ltep |~\theta(\sysp)\leq \alpha}$. Apparently, the set $\mathcal{P}(\alpha)$ is a \emph{cone}, namely, $k \sysp \in \mathcal{P}(\alpha)$ for all $\sysp \in \mathcal{P}(\alpha)$ and $k>0$. In addition, the set $\mathcal{P}(\alpha)$ is \emph{closed} under the feedback map $\sysg$. This can be concluded from Proposition~\ref{prop:large_angle} as a notable special case.
\begin{cor}\label{prop:small_angle}
For $\alpha\in \interval[open right]{0}{\pi/2}$ and $\sysp, \sysc \in \mathcal{P}(\alpha)$, it holds that $\sysg \in \mathcal{P}(\alpha)$.
\end{cor}
\pf
For a well-posed feedback system $\gof|_{e_2=0}$ in Fig.~\ref{fig:feedback}, by hypothesis, the small angle condition (\ref{eq:spt}) is satisfied and thus $\boldsymbol{G}=e_1\mapsto y_1$ is stable according to Theorem~\ref{thm:spt}. By using $\max\bbkt{\theta(\sysp),\theta(\sysc)}\leq \alpha$ in Proposition~\ref{prop:large_angle}, we obtain that $\sysg \in \mathcal{P}(\alpha)$.
 \hfill $\square$
\endpf
 
Proposition~\ref{prop:large_angle} and Corollary~\ref{prop:small_angle} also hold for a well-posed feedback system $\gof|_{e_1=0}$.

For systems $\sysp_1$ and $\sysp_2$, a parallel interconnected system is defined to be $\sysp_1+\sysp_2$. The next proposition links $\theta(\sysp_1)$ and $\theta(\sysp_2)$ to $\theta(\sysp_1+\sysp_2)$.

\begin{prop}\label{prop:parallel}
 For $\sysp_1, \sysp_2$ with $\theta(\sysp_1), \theta(\sysp_2)\leq \pi/2$, it holds that $\theta(\sysp_1+\sysp_2)\leq \max\bbkt{\theta(\sysp_1), \theta(\sysp_2)}$. 
\end{prop}
\pf
 Let $u=u_1=u_2$ and $y=y_1+y_2$. WLOG, assume that $\theta(\sysp_1)\geq \theta(\sysp_2)$. By hypothesis, we have
 \begin{align*}
 \ininf{u}{y}
 &\geq \cos\theta(\sysp_1)\norm{u_1}_2\norm{y_1}_2+\cos\theta(\sysp_2)\norm{u_2}_2\norm{y_2}_2\\
 &\geq \cos\theta(\sysp_1)\norm{u}_2\rbkt{\norm{y_1}_2+\norm{y_2}_2}\\
 &\geq \cos\theta(\sysp_1)\norm{u}_2\norm{y}_2,
 \end{align*}
 which implies that $\theta(\sysp_1+\sysp_2)\leq \theta(\sysp_1)$. We thus conclude that $\theta(\sysp_1+\sysp_2)\leq \max\bbkt{\theta(\sysp_1), \theta(\sysp_2)}$. \hfill $\square$
\endpf

The singular angles of subsystems in Corollary~\ref{prop:small_angle} and Proposition~ \ref{prop:parallel} are restricted to be no greater than $\pi/2$. These subsystems are thus passive. Comparing with the feedback and parallel interconnection results of passive systems, e.g., see \cite[Sec.~3.2] {Sepulchre:97}, we have provided quantitative descriptions of the interconnection results based on the singular angle. In light of Propositions~\ref{prop:productPC}, \ref{prop:large_angle} and \ref{prop:parallel} and Corollary~\ref{prop:small_angle}, the singular angle of a nonlinear network consisting of subsystems through suitable cascaded, parallel and feedback interconnections can be estimated by the singular angles of the subsystems. 

\section{Extensions and Further Discussions}\label{sec:09}
The purpose of this section is threefold. The first is dedicated to a notable extension and an $\mathcal{L}_{2e}$-variation of the singular angle. We come up with the generalized singular angle via the use of multipliers, which can reduce the conservatism of Theorem~\ref{thm:spt}. The second is to draw a comparison between the singular angle and the recent system phase \cite{Chen:20j}, since both of the works aim at analyzing nonlinear systems from an angular or phasic viewpoint. The third is to discuss the computational issue of the singular angle with possible methods for future works.

\subsection{The Generalized Singular Angle}
Over the past half century, the multiplier approach has been widely adopted in feedback system analysis. The key idea behind the method is to leverage suitable complementary multipliers for a feedback system so that the system may satisfy some ``weighted'' stability conditions. This certainly reduces the conservatism of feedback system analysis. Representative works including the multiplier theorem \cite{Zames:68}, $(Q, S, R)$-dissipativity theory\cite{Hill:80} and IQCs theory \cite{Megretski:97, Khong:21}. These works inspire us to incorporate the use of multipliers into system singular angles.

Let $\mathcal{B}(\mathcal{L}_2(-\infty, \infty))$ denote the set of linear bounded invertible operators mapping $\lt(-\infty, \infty)$ to $\lt(-\infty, \infty)$. Consider multipliers $\sysm_1, \sysm_2\in \mathcal{B}(\mathcal{L}_2(-\infty, \infty))$. For $u, v\in \ltp$, define the generalized angle $\theta_{\sysm_1, \sysm_2}(u,v) \in \interval{0}{\pi}$ between $u$ and $v$ with respect to $\sysm_1$ and $\sysm_2$ by the formula
 \be\label{eq: generalized_angle}
 \cos\theta_{\sysm_1,\sysm_2}(u, v)\coloneqq\dfrac{\ininf{\sysm_1 u}{\sysm_2 v}}{\norm{\sysm_1 u}_2\norm{\sysm_2 v}_2},
 \ee
if $u,v \in \ltp\setminus \{0\}$, and $\theta_{\sysm_1,\sysm_2}(u, v)\coloneqq 0$, if $u=0$ or $v=0$. Here we slightly abuse the notation $\ininf{\cdot}{\cdot}$ and $\|\cdot\|_2$ for elements in $\lt(-\infty, \infty)$ rather than causal $\lt$. A generalization of Lemma~\ref{lem:signal_triangle_ineq} which involves the multipliers is as follows.
 \begin{lem}\label{lem:G_signal_triangle_ineq}
 Consider $\sysm_1, \sysm_2, \sysm_3\in \mathcal{B}(\mathcal{L}_2(-\infty, \infty))$.
 For all $u, w\in \ltp$ and $0\neq v \in \ltp$, it holds that
 $\theta_{\sysm_1, \sysm_3}(u,w)\leq \theta_{\sysm_1, \sysm_2}(u, v)+\theta_{\sysm_2, \sysm_3}(v,w)$.
 \end{lem}
 \pf
For $u ,w\in \ltp$ and $v\in \ltp\setminus\{0\}$, denote $\bar{u}\coloneqq \sysm_1 u$, $\bar{w}\coloneqq \sysm_3 w$ and $\bar{v}\coloneqq \sysm_2 v$. Since $\sysm_1, \sysm_2$ and $\sysm_3$ are bounded on $\lt(-\infty, \infty)$, it holds that $\bar{u},\bar{w}\in \lt(-\infty, \infty)$ and $\bar{v} \in \lt(-\infty, \infty)\setminus\{0\}$. Following the same reasoning as in the proof of~\cite[Lem.~3.3-1]{Gustafson:97}, we obtain $\theta(\bar{u},\bar{w}) \leq \theta(\bar{u},\bar{v})+\theta(\bar{v},\bar{w})$,
where the definition of the angle $\theta(\cdot, \cdot)$ between $\bar{u}$, $\bar{v}$, $\bar{w}$ is modified to fit the space $\lt(-\infty, \infty)$. Due to~\eqref{eq: generalized_angle}, this gives that $
 \theta_{\sysm_1, \sysm_3}(u,w)\leq \theta_{\sysm_1, \sysm_2}(u, v)+\theta_{\sysm_2, \sysm_3}(v,w)$. \hfill $\square$
 \endpf
 
Given a causal stable system $\sysp\colon \ltep\to\ltep$ and multipliers $\sysm_1, \sysm_2\in \mathcal{B}(\mathcal{L}_2(-\infty, \infty))$, we define the \textit{generalized singular angle} $\theta_{\sysm_1,\sysm_2}(\sysp)\in \interval{0}{\pi}$ with respect to $\sysm_1$ and $\sysm_2$ via
 \bex
 \displaystyle \cos\theta_{\sysm_1,\sysm_2}(\sysp)\coloneqq \displaystyle\inf_{\substack{ 0\neq u \in \ltp,\\{\sysp u\neq 0}}} \dfrac{\ininf{\sysm_1 u}{\sysm_2\sysp u}}{\norm{\sysm_1 u}_2\norm{\sysm_2\sysp u}_2}.
 \eex
The conservatism of Theorem~\ref{thm:spt} can be reduced to an extent by the use of appropriate multipliers. We state the following generalized version of Theorem~\ref{thm:spt}.
\begin{thm}\label{thm:gspt}
For causal stable $\sysp$ and $\sysc$, the well-posed $\gof|_{e_2=0}$ is stable if there exist a unitary multiplier $\sysm_1\in \mathcal{B}(\mathcal{L}_2(-\infty, \infty))$ and a multiplier $\sysm_2\in \mathcal{B}(\mathcal{L}_2(-\infty, \infty))$ so that
$\theta_{\sysm_1, \sysm_2}(\sysp) + \theta_{\sysm_2, \sysm_1}(\sysc) < \pi$.
\end{thm}
\pf
The proof can be shown by following an analogous procedure to the proof of Theorem~\ref{thm:spt}, except for the following differences. Note that $\sysm_1$ and $\sysm_2$ are linear, bounded and invertible. By hypothesis, we have $\cos\sbkt{\theta_{\sysm_1, \sysm_2}(\sysp) + \theta_{\sysm_2, \sysm_1}(\sysc)} > -1$,
which gives that $\cos\sbkt{\theta_{\sysm_1, \sysm_2}(u_{1}, y_{1}) + \theta_{\sysm_2, \sysm_1}(u_{2}, y_{2})} > -1$
for all $u_1, u_2\in\ltp\setminus\{0\}$. Applying Lemma~\ref{lem:G_signal_triangle_ineq} and noting $y_1=u_2$ yield that 
 \begin{align*}
 -1&<\cos\sbkt{\theta_{\sysm_1, \sysm_2}(u_{1}, y_{1}) + \theta_{\sysm_2, \sysm_1}(u_{2}, y_{2})}\\
 &\leq \cos\theta_{\sysm_1, \sysm_1}(u_1, y_2) \leq 1
 \end{align*}
for all $u_1, u_2\in\ltp\setminus\{0\}$. Thus, $\theta_{\sysm_1, \sysm_1}(u_1, y_2)\in \interval[open right]{0}{\pi}$. Since $\sysm_1$ is unitary, i.e. $\sysm_1^{*}=\sysm_1^{-1}$, by \eqref{eq: generalized_angle} we obtain $ \theta_{\sysm_1, \sysm_1}(u_1, y_2)=\theta(u_1, y_2) \in \interval[open right]{0}{\pi}$. Note that $\theta_{\sysm_1, \sysm_1}(\cdot, \cdot)$ involves signals in $\lt(-\infty, \infty)$, while $\theta(\cdot, \cdot)$ contains only signals in $\lt$. Following~\eqref{eq:sysineq}, the proof is then completed with a homotopy argument by the same reasoning as in the proof of Theorem~\ref{thm:spt}.
\hfill $\square$ 
\endpf

The next corollary is directly obtained by specifying $\sysm_1=\sysi$ in Theorem~\ref{thm:gspt}, which is more relevant to the classical multiplier setup. 
\begin{cor}\label{coro:multiplier_spt}
 For causal stable $\sysp$ and $\sysc$, the well-posed $\gof|_{e_2=0}$ is stable if there exists a multiplier $\sysm\in \mathcal{B}(\mathcal{L}_2(-\infty, \infty))$ such that $\theta_{\sysi, \sysm}(\sysp) + \theta_{\sysm, \sysi}(\sysc) < \pi$.
\end{cor}

In the setup above, the multipliers $\sysm$ and $\sysm^*$ appear as a conjugate pair for simultaneously weighting $\sysp$ and $\sysc$. The singular angle is applicable to a network structure in Fig.~\ref{fig:cyc}. How to formulate a statement with compatible multipliers $\sysm_1, \sysm_2, \ldots, \sysm_N$ for subsystems $\sysp_1, \sysp_2, \ldots, \sysp_N$ remains unclear and deserves further investigation.

\subsection{The $\mathcal{L}_{2e}$ Singular Angle}
System properties can be defined on different signal spaces, e.g., $\ltp$ or $\ltep$. We have seen in Section~\ref{sec:02} that for causal stable nonlinear systems, the definition of $\lt$-gain is equivalent to that of $\mathcal{L}_{2e}$-gain, and an analogous equivalence applies to the definition of passivity. This motivates us to consider the system singular angle on $\ltep$ and explore the existence of a similar equivalence.

For a causal stable system $\sysp \coloneqq u\mapsto y\colon \ltep \to \ltep$, the \textit{$\mathcal{L}_{2e}$ singular angle} $\theta_e(\sysp)\in \interval{0}{\pi}$ is defined by 
\be\label{eq:extended_angle}
 \theta_e(\sysp) \coloneqq \displaystyle\sup_{\substack{ u \in \ltep, y=\syspu, ~T> 0,\\ {\norm{u_T}_2\neq 0},{\norm{ y_T}_2\neq 0}}} \arccos\frac{\ininf{u_T}{y_T}}{\norm{u_T}_2\norm{y_T}_2}.
\ee

\begin{prop}\label{prop:extended_angle}
 For causal stable $\sysp$, we have $
 \theta(\sysp)\leq \theta_e(\sysp)$.
 In addition, if $\theta(\sysp)\in \interval{0}{\frac{\pi}{2}}$, then $\theta(\sysp)=\theta_e(\sysp)$.
\end{prop}
\pf
 The case $\theta(\sysp)\leq \theta_e(\sysp)$ is proved by taking $T\rightarrow \infty$ in (\ref{eq:extended_angle}). We next show that $\theta_e(\sysp)\leq \theta(\sysp)$ when $\theta(\sysp)\in \interval{0}{\pi/2}$. Since $\sysp$ is causal, then $(\sysp u)_T=(\sysp u_T)_T$ holds for $T>0$. By definition~\eqref{eq:extended_angle}, we have
 \begin{align*}
 \cos\theta_e(\sysp) &= \inf_{\substack{ 0\neq u_T \in \ltp, T>0,{\sysp u_T\neq 0}}}\frac{\ininf{u_T}{(\sysp u_T)_T}}{\norm{u_T}_2\norm{(\sysp u_T)_T}_2} \\
 &=\inf_{\substack{ 0\neq u_T \in \ltp, T>0, {\sysp u_T\neq 0}}}\frac{\ininf{u_T}{\sysp u_T}}{\norm{u_T}_2\norm{(\sysp u_T)_T}_2}
 \\
 &\geq\inf_{\substack{ 0\neq u_T \in \ltp, T>0 \\{\sysp u_T\neq 0}}}\frac{\ininf{u_T}{\sysp u_T}}{\norm{u_T}_2\norm{\sysp u_T}_2}=\cos\theta(\sysp), 
\end{align*}
where the fact $\norm{(\sysp u_T)_T}_2\leq \norm{\sysp u_T}_2$ is due to that $\norm{f_T}_2$ is an increasing function of $T$ and $\norm{f}_2=\lim_{T\rightarrow \infty} \norm{f_T}_2$. We conclude $\theta_e(\sysp)\leq \theta(\sysp)$. \hfill $\square$  
\endpf
We have shown that $\theta(\sysp)$ and $\theta_e(\sysp)$ are generally unequal. When $\theta(\sysp)\in \interval[open left]{\pi/2}{\pi}$, the sign of $\ininf{u_T}{\sysp u_T}$ in the proof of Proposition~\ref{prop:extended_angle} can be \emph{indefinite}. Therefore, one cannot deduce the equivalence between $\theta(\sysp)$ and $\theta_e(\sysp)$. Equipped with the {$\mathcal{L}_{2e}$ singular angle}, we immediately obtain the following corollary from Theorem~\ref{thm:spt}.
\begin{cor}
For causal stable $\sysp$ and $\sysc$, the well-posed $\gof|_{e_2=0}$ is stable if
$\theta_e(\sysp) + \theta_e(\sysc) < \pi$.
\end{cor}
 
\begin{table*}[htb]
\begin{center}
\begin{tabular}{c c c}
\toprule[1pt] Property & Singular angle $\theta(\sysp)$ & Phase $\Phi(\sysp)=\interval{\underline{\phi}(\sysp)}{\overline{\phi}(\sysp)}$\\
 \specialrule{0em}{1pt}{1pt}
\hline
 \specialrule{0em}{2pt}{2pt}
Value & A quantity $\theta(\sysp)\in \interval{0}{\pi}$ & A $\pi$-length interval $\Phi(\sysp)\subset \interval{-3\pi/2}{3\pi/2}$ \\
 \specialrule{0em}{1pt}{1pt}
Applicable scope & All systems & Semi-sectorial systems\\
 \specialrule{0em}{1pt}{1pt}
{Feedback stability condition} & $\theta(\sysp)+\theta(\sysc)<\pi$ & \makecell{$\overline{\phi}(\sysp) + \overline{\phi}(\sysc) < \pi$,\\
$\underline{\phi}(\sysp) + \underline{\phi}(\sysc) > -\pi$}\\
 \specialrule{0em}{1pt}{1pt}
 Cascaded interconnection & $\theta(\sysp_2\sysp_1)\leq \theta(\sysp_1)+\theta(\sysp_2)$ & $/$ \\
 \specialrule{0em}{1pt}{1pt}
Parallel interconnection & $/$ & \makecell{$\Phi(\sysp_1),\Phi(\sysp_2)\subset \interval{\alpha}{\beta}$ $\Rightarrow$ $\Phi(\sysp_1+\sysp_2)\subset \interval{\alpha}{\beta}$} \\
 \specialrule{0em}{1pt}{1pt}
Feedback interconnection & $\theta(\boldsymbol{G})\leq \max\bbkt{\theta(\sysp),\theta(\sysc)}$ & $ \Phi(\boldsymbol{G}) \subset \interval {\min\bbkt{\underline{\phi}(\sysp),-\overline{\phi}(\sysc)}}{\max\bbkt{\overline{\phi}(\sysp),-\underline{\phi}(\sysc)}}$ \\
 \specialrule{0em}{1pt}{1pt}
Time domain vs frequency domain & Nonequivalent & Equivalent \\
 \specialrule{0em}{1pt}{1pt}
Interpretation & \makecell{The rotating effect from \\ system input to output signals} & \makecell{The tradeoff between \\the real energy and reactive energy} \\
\bottomrule[1pt]
\end{tabular}
\end{center}
\caption{A comparison between the system singular angle and system phase}\label{tab:angle_phase}
\end{table*}

\subsection{A Comparison with the Recent Nonlinear Phase}\label{subsec:phase}
A practical nonlinear system can only accept and generate real-valued signals. To define a phase notion in nonlinear systems, one may have two feasible paths. The first is what we have done in this paper, inspired by the Euclidean space angle between vectors. The second is to introduce complex elements to nonlinear systems on the grounds that phases are naturally defined for complex numbers. To the second end, the authors of \cite{Chen:20j} proposed the notion of the nonlinear system phase through complexifying real-valued signals using the Hilbert transform. Concretely, for a causal stable system $\sysp$, the \textit{angular numerical range} of $\sysp$ is defined to be \cite{Chen:20j}: \bex
W(\sysp) \coloneqq \bbkt{ \ininf{0.5(u+j\sysh u)}{\syspu} \in \mathbb{C}~|~ u \in \ltp }, \eex
where $\sysh$ denotes the Hilbert transform which is frequently used in signal processing.
Such a system $\sysp$ is said to be \textit{semi-sectorial} if $W(\sysp)$ is contained in a closed complex half-plane. Then, for semi-sectorial systems, the \textit{phase} of $\sysp$, denoted by ${\Phi}(\sysp)$, is defined to be the \textit{phase sector} ${\Phi}(\sysp) \coloneqq \interval{\underline{\phi}(\sysp)}{\overline{\phi}(\sysp)}$,
where ${\underline{\phi}(\sysp)}$ and ${\overline{\phi}(\sysp)}$ are called the phase infimum and phase supremum of $\sysp$, respectively, and are defined by
\bex
{\underline{\phi}}(\sysp) \coloneqq \inf_{0\neq z \in W(\sysp)} {\angle z}\quad \text{and}\quad
{\overline{\phi}}(\sysp) \coloneqq \sup_{0\neq z \in W(\sysp)} {\angle z}.
\eex
The system singular angle and the system phase are in general different, though both extend the classical phase notion. This can be easily understood from a matrix perspective. On the one hand, consider a positive definite matrix $A=\stbt{1}{0}{0}{2}$ which is semi-sectorial. Its phase $\Phi(A)=\interval{0}{0}$ is zero \cite{Chen:21}, while its singular angle $\theta(A)= \frac{19.4712\pi}{180}$ is nonzero and seems conservative as one might expect its angle being zero as well. On the other hand, for those matrices which are not semi-sectorial, their phases are undefined. Their singular angles however can still be computed. We refer the reader to \cite{Chen:20j} for more details of the system phase and we end this section by comparing the system singular angle with the system phase with the summary in Table~\ref{tab:angle_phase}. 
 
\subsection{Discussions on the Computational Issue}
We briefly discuss the difficulties and possible methods toward computation of the singular angle for matrices and systems. For matrices, the computation of the singular angle can be addressed from two perspectives. First, formula \eqref{eq:matrix_angle} can be linked to the notion of normalized numerical range $\{ \frac{x^* A x}{\abs{x}\abs{Ax}}|~0\neq x\in \cn, Ax\neq 0\}$. A systematic approach to generating the normalized numerical range was proposed in \cite{Auzinger:03} analogously to the numerical range \cite[Sec.~5.6]{Gustafson:97}. Second, formula \eqref{eq:matrix_angle} leads to nonlinear fractional programming that may be converted into parametric programming and partially solved using Dinkelbach's algorithm \cite[Ch.~4]{Stancu:12}. How to solve the problem efficiently is nontrivial and is our ongoing research. As for nonlinear systems, computing the singular angle is generally difficult beyond suitable estimation, similarly to other input-output quantities. One promising way is to leverage state-space dissipativity theory \cite{Hill:80}, which seeks for the solutions to algebraic inequalities characterized by the notion of supply rate $s(u, y)(t)$ as an abstraction of physical power, i.e., the nonlinear KYP lemma \cite[Th.~17]{Hill:80}, where $y=\sysp u$. For instance, an $\lt$-gain $\gamma$ in \eqref{eq:l2gain} can be described by the \emph{static quadratic} supply rate $s(u, y)(t)=\gamma^2 |u(t)|^2 - |y(t)|^2$ and passivity indices $\rho, \nu$ in \eqref{eq:VSP} by $s(u, y)(t)=u(t)^\top y(t) - \rho |u(t)|^2 -\nu \abs{y(t)}^2$. Exploring an appropriate supply rate that precisely describes the singular angle thus becomes the \emph{nucleus} of resolving the computation problem. Nevertheless, we have to acknowledge that it is highly nontrivial to extract a supply rate from definition~\eqref{eq:nonlinear_angle} (possibly a dynamic one \cite{Khong:22}). The integral-product term $\norm{u}_2\norm{y}_2$ is \emph{non-quadratic} and how to disclose a single integrand from the term to obtain the supply rate is unclear yet. Moreover, notice that there is a minor gap between the singular angle and its $\mathcal{L}_{2e}$ version as stated earlier. Once a KYP-type lemma for the $\mathcal{L}_{2e}$ singular angle becomes available, we need to address the gap to obtain the singular angle. We leave the state-space framework of the singular angle as future research.

\section{A Simulation Example}\label{sec:simulation}
We provide a simulation example for demonstrating the small angle theorem. Consider a cyclic system in Fig.~\ref{fig:cyc} with subsystems $\sysp_i$ for $i=1, 2, 3, 4$. Let $\sysp_1$ be a stable LTI system with $P_1(s)=\tau \frac{(s+14)(s^2+2s+0.1)}{(s^2+s+2)(s^2+5s+20)}$, where $\tau >0$ is arbitrary so that $\norm{\sysp_1}$ can be made arbitrarily large, whereby the small gain theorem is inapplicable. Let $\sysp_3$ have the following state-space representation:
\begin{align*}
 \dot{x}_1(t)&=x_2(t)\\
 \dot{x}_2(t)&=-\rbkt{x_1(t)}^3-0.3x_2(t)+1.4u_3(t)\\
 y_3(t)&=x_2(t)+u_3(t)
 \end{align*}
 with $\stbo{x_1(0)}{x_2(0)}=\stbo{0}{0}$. Let $\sysp_2$ and $\sysp_4$ both be logarithmic quantizers with different quantization densities $\tilde{\rho}_2={1}/{3}$ and $\tilde{\rho}_4=0.8$; i.e., each one is a nonlinearity $\sysn$ defined by \eqref{eq:sector}, with $h\colon \mathbb{R}\rightarrow\mathbb{R}$ satisfying
 \be\label{eq:quantizer}
 h(x) = \left\{
 \begin{array}{ll}
 \tilde{\rho}^i, &\text{if}~x \in \interval[open left]{\frac{1+\tilde{\rho}}{2}\tilde{\rho}^i}{ \frac{1+\tilde{\rho}}{2\tilde{\rho}}\tilde{\rho}^i},~i=0, \pm1, \ldots\\
 0,&\text{if}~x=0,\\
 -h(-x),&\text{if}~x<0,
 \end{array}
 \right.
 \ee
 where $\tilde{\rho} \in \interval[open]{0}{1}$ represents the quantization density. Such a feedback system subject to both input and output quantization has been widely studied, e.g., see \cite{Coutin:10}.  

First, note that $\angle P_1(\jw)\subset \interval{\frac{-118.32\pi}{180}}{\frac{77.91\pi}{180}}$ for all $\omega\in \interval{0}{\infty}$ and thus $\theta(\sysp_1)= \frac{118.32\pi}{180}$ by Proposition~\ref{prop:SISO}. Second, one can verify that $\sysp_3$ is stable and very strictly passive with passivity indices $ \nu=0.7$ and $ \rho= 0.3$. By Proposition~\ref{prop:angle_VSP}, we have $\theta(\sysp_3)\leq \frac{23.58\pi}{180}$. Third, it is known \cite{Fu:05_2} that $\sysn$ defined with \eqref{eq:quantizer} belongs to a nonlinearity sector from ${\frac{2\tilde{\rho}}{1+\tilde{\rho}}}$ to ${\frac{2}{1+\tilde{\rho}}}$. It then follows from Corollary~\ref{prop:nonlinearity} that $\theta(\sysn)\leq \arccos \frac{2\sqrt{\tilde{\rho}}}{1+\tilde{\rho}}$ and the quantization densities $\tilde{\rho}_2={1}/{3}$ and $\tilde{\rho}_4=0.8$ give the singular angles $\theta(\sysp_2)\leq {\frac{\pi}{6}}$ and $\theta(\sysp_4)\leq {\frac{6.38\pi}{180}}$. Therefore, it holds that $\theta(\sysp_1) +\theta(\sysp_2) +\theta(\sysp_3)+\theta(\sysp_4) <\pi$. By Theorem~\ref{thm:spt} and Corollary~\ref{thm:cyclic}, the cyclic system is stable for all $\tau>0$. To the best of our knowledge, such a conclusion for four subsystems may not be directly drawn from existing IQCs/dissipativity stability results.

\begin{figure}[htb]
 \centering
 \includegraphics[width=2.5in, trim={0cm 0cm 0cm 0cm}, clip]{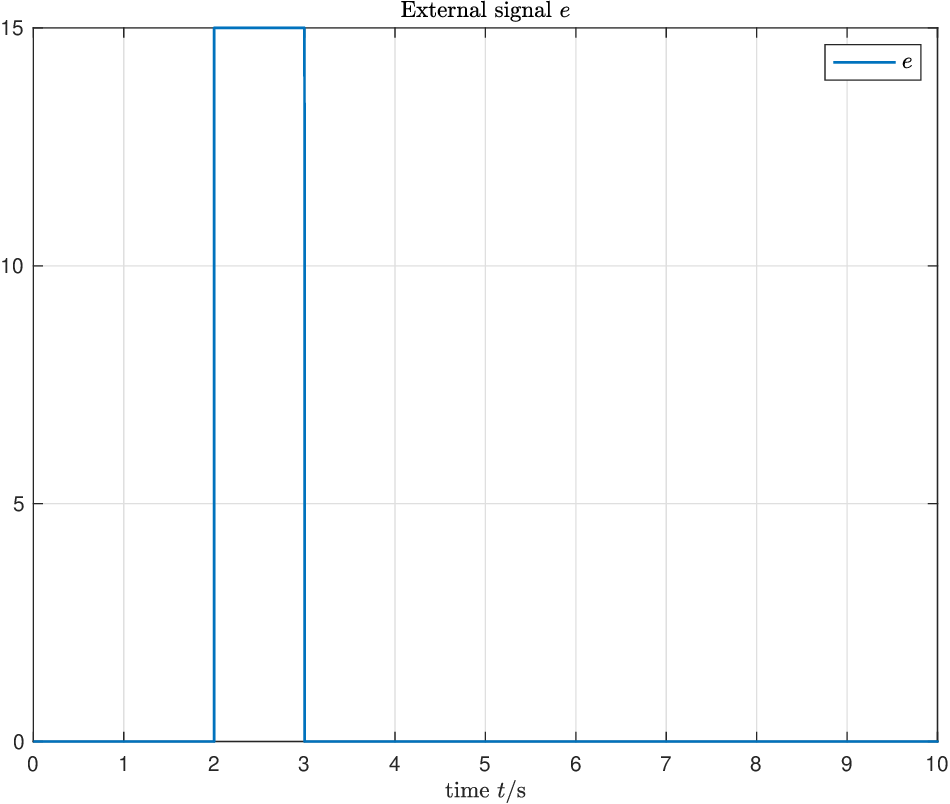}
 \includegraphics[width=2.5in]{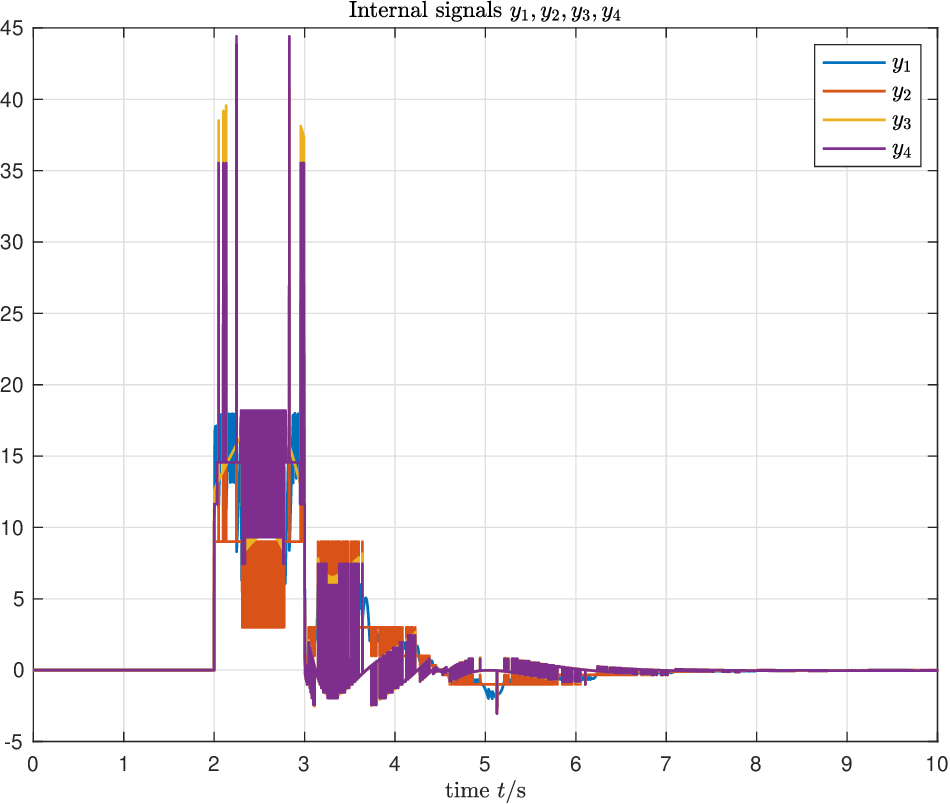}
 \caption{The external signal $e$ (top) and internal signals $y_1$, $y_2$, $y_3$, $y_4$ (bottom) of the cyclic system when $\tau=500$. The external signal consists of one rectangular pulse and all the internal signals converge to zero in about eight seconds.}
\label{fig:signals}
\end{figure}

 As depicted in Fig.~\ref{fig:signals}, to test the stability, we adopt a rectangular pulse to generate the external signal $e$ to stimulate the cyclic system when $\tau = 500$. The corresponding responses of the internal signals $y_i$ for $i=1, 2, 3, 4$ converge to zero within eight seconds.  

\section{Conclusion}\label{sec:10}
In this paper, we propose an angle notion, the singular angle, for nonlinear systems from an input-output perspective. The system singular angle plays the role of a counterpart to the system $\lt$-gain and can quantify the system passivity. Subsequently, we establish the main result, a nonlinear small angle theorem, for feedback stability analysis, which is a successful realization of the Zames' envision in 1966. We also investigate the singular angles of appropriate feedback, parallel or cascaded interconnected systems. Finally, we develop a generalization of the small angle theorem for reducing the conservatism in feedback stability analysis. 

Ongoing research directions include computational issues and controller synthesis problems of the system singular angle. An angle study of linear stochastic systems is provided in \cite{Zhao:23_Angle}. It is hoped that this paper provides a new starting point for bringing the recent renaissance of the classical system phase notion into nonlinear systems. 

\appendix
\section{Appendix}
{\bfseries PROOF of Proposition~\ref{prop:SISO}.}
 We first show \ref{item: SISO_angle}: ${\theta_\infty}(P)\leq \theta(\sysp)$.
 Suppose ${\theta}_\infty(P)$ defined in \eqref{eq:siso_LTI_sphase} is attained at some $\omega_0\in \interval{-\infty}{\infty}$. Let $u_0\in \ltp$ with its $\widehat{u_0}$ chosen such that 
 \be\label{eq: unit_signal}
 \hspace{-3mm}\abs{\widehat{u_0}({\jw})}=\left\{
 \begin{array}{l}
 c\quad \text{if}~\abs{\omega+\omega_0}<\epsilon~ \text{or}~\abs{\omega-\omega_0}<\epsilon \\
 0\quad \text{otherwise},
 \end{array}
 \right.
 \ee
where $\epsilon$ is a small positive number and $c$ is chosen so that $\widehat{u_0}$ has a unit 2-norm, i.e., $c=\sqrt{{\pi}/({2\epsilon})}$. Using the Plancherel's theorem, we have that
\begin{align*}
 &\frac{\ininf{u_0}{\sysp u_0}}{\norm{u_0}_2\norm{\sysp u_0}_2}=\frac{\ininf{\widehat{u_0}}{\widehat{\sysp u_0}}}{\norm{\widehat{u_0}}_2\norm{\widehat{\sysp u_0}}_2}\\
 =&\frac{\frac{1}{2\pi}\int_{-\infty}^{\infty}{\rep \rbkt{P(\jw)} \abs{\widehat{u_0}(j\omega)}^2}\,\mathrm{d}\omega}{\sqrt{\frac{1}{2\pi}\int_{-\infty}^{\infty}{\abs{P(j\omega)}^2
 \abs{\widehat{u_0}(j\omega)}^2}\,\mathrm{d}\omega}}\rightarrow \frac{\rep \rbkt{P(j\omega_0)}}{\abs{P(j\omega_0)}}
 \end{align*}
 as $\epsilon \rightarrow 0$. It follows that
 \bex
 \inf_{\substack{ 0\neq u \in \ltp,\\{\sysp u\neq 0}}} \dfrac{\ininf{u}{\sysp u}}{\norm{u}_2\norm{\sysp u}_2} \leq \inf_{\substack{ \omega\in \interval{-\infty}{\infty},\\{P(\jw)\neq 0}}} \frac{\rep \rbkt{P(\jw)}}{\abs{P(\jw)}}
 \eex
 and ${\theta_\infty}(P) \leq \theta(\sysp)$. Next, we show \ref{item: SISO_angle_2}: $\theta(\sysp) \leq {\theta_\infty}(P)$ when ${\theta_\infty}(P)\in \interval[open left]{\pi/2}{\pi}$. Since $ \cos{\theta_\infty}(P) \in \interval[open right]{-1}{0}$ and for all $\omega \in \interval{-\infty}{\infty}$, it holds that $\rep \rbkt{P(\jw)} \geq \cos{\theta_\infty}(P) \abs{P(\jw)}$.
 Then, for all $u\in \ltp\setminus\{0\}$, we have 
 \begin{multline}\label{eq:prop_SISO_0}
 \hat{u}(\jw)^*\rep\rbkt{P(\jw)} \hat{u}(\jw) \geq \cos{\theta_\infty}(P) \abs{P(\jw)}\abs{\hat{u}(\jw)}^2\\
 =\cos{\theta_\infty}(P) \abs{P(\jw)\hat{u}(\jw)}\abs{\hat{u}(\jw)}.
 \end{multline}
 Integrating both sides of the above inequality gives
 \begin{align}\label{eq:prop_SISO}
 \hspace{-1em} &\ininf{\hat{u}}{P\hat{u}}\geq \frac{ \cos{\theta_\infty}(P)}{2\pi}\int_{-\infty}^{\infty}\abs{P(\jw)\hat{u}(\jw)}\abs{\hat{u}(\jw)}\,\mathrm{d}\omega\notag\\
 \hspace{-0.7em} =&\cos{\theta_\infty}(P)\ininf{\abs{\hat{u}}}{\abs{P\hat{u}}}\geq \cos{\theta_\infty}(P) \norm{\hat{u}}_2 \norm{P\hat{u}}_2, 
 \end{align}
 where the first inequality follows from integrating \eqref{eq:prop_SISO_0} and the conjugate symmetry $P(-\jw) = P(\jw)^*$ and the last inequality uses the Cauchy-Schwarz
 inequalities and $\cos{\theta_\infty}(P)<0$. Rearranging (\ref{eq:prop_SISO}) and by the Plancherel's theorem yield $\frac{\ininf{u}{{\sysp u}}}{\norm{{u}}_2\norm{{\sysp u}}_2} \geq \cos{\theta_\infty}(P)$.
 Hence, we obtain 
 $\cos\theta(\sysp)=\inf_{\substack{ 0\neq u \in \ltp,\\{\sysp u\neq 0}}} \frac{\ininf{u}{\sysp u}}{\norm{u}_2\norm{\sysp u}_2} \geq \cos{\theta_\infty}(P)$
 and conclude $\theta(\sysp)\leq {\theta_\infty}(P)$. \hfill $\square$

We next provide a statement that complements Proposition~\ref{prop:SISO}. When $\theta_\infty(P)$ defined in \eqref{eq:siso_LTI_sphase} is contained in $\interval[open right]{0}{\pi/2}$, we will show that the strict inequality ${\theta_\infty}(P)< \theta(\sysp)$ holds via constructing a numerical example. Construct a signal $f\in \ltp$ with its $\hat{f}$ chosen such that
\bex
\abs{\hat{f}({\jw})}=\left\{
\begin{array}{l}
 c_0\quad \text{if}~\abs{\omega+\omega_0}<\epsilon~ \text{or}~\abs{\omega-\omega_0}<\epsilon, \\
 c_1\quad \text{if}~\abs{\omega+\omega_1}<\epsilon~ \text{or}~\abs{\omega-\omega_1}<\epsilon, \\
 0\quad \text{otherwise},
\end{array}
\right.
\eex
where $\epsilon$ is a small positive number, $\omega_0, \omega_1\geq 0$, and $c_0, c_1>0$ are chosen so that $\hat{f}$ has a unit 2-norm, i.e.,
\be\label{eq:appendix_B_01}
c_0^2+c_1^2={\pi}/\rbkt{2\epsilon}.
\ee
In light of definition (\ref{eq:nonlinear_angle}), using this particular $f$ yields
\begin{align}\label{eq:appendix_B_02}
 &\cos\theta(\sysp)\leq \dfrac{\ininf{f}{\sysp f}}{\norm{f}_2\norm{\sysp f}_2}={\ininf{\hat{f}}{\widehat{\sysp f}}}\big/{\norm{\widehat{\sysp f}}_2} \notag\\
 =&~\frac{\frac{2\epsilon}{\pi}\rbkt{\rep P(\jw_0)c_0^2 + \rep P(\jw_1)c_1^2}}{\sqrt{\frac{2\epsilon}{\pi} \rbkt{\abs{P(\jw_0)}^2c_0^2+\abs{P(\jw_1)}^2c_1^2}}} \notag\\
 =&~\frac{\frac{2\epsilon}{\pi}c_0^2 \rep P(\jw_0) + \rbkt{1-\frac{2\epsilon}{\pi}c_0^2}\rep P(\jw_1) }{\sqrt{\frac{2\epsilon}{\pi}c_0^2\abs{P(\jw_0)}^2+ \rbkt{1-\frac{2\epsilon}{\pi}}\abs{P(\jw_1)}^2 }},
 \end{align}
 where the second-last equality follows from~\eqref{eq:appendix_B_01}.
 Assume that $\theta_\infty(P)$ is attained at $\omega_0$, namely, $\cos\theta_\infty(P)=\frac{\rep \rbkt{P(\jw_0)}}{\abs{P(\jw_0)}}$. Denote $P(\jw_0)=z_0\coloneqq a_0+jb_0 \in \mathbb{C}$ and $P(\jw_1)=z_1\coloneqq a_1+jb_1\in \mathbb{C}$ with $a_0, a_1>0$. Since $\theta_\infty(P)$ is attained at $\omega_0$, then $z_1, z_2$ should further satisfy $\abs{\angle z_0} \geq \abs{\angle z_1}$. Denote $\tau\coloneqq 2\epsilon c_0^2 /\pi$. It follows from~\eqref{eq:appendix_B_01} that $c_0^2<\pi/(2\epsilon)$ and $0<2\epsilon c_0^2/\pi<1$. Thus, $\tau\in\interval[open]{0}{1}$. We rewrite \eqref{eq:appendix_B_02} as
 $\cos\theta(\sysp)\leq \textstyle \frac{\tau a_0+(1-\tau)a_1}{\sqrt{\tau (a_0^2+b_0^2)+(1-\tau)(a_1^2+b_1^2)}}$ 
and $\cos\theta_\infty(P)=\frac{a_0}{\sqrt{a_0^2+b_0^2}}$. We aim to find feasible $\tau, z_0$ and $z_1$ for some systems $\sysp$ to show that $\theta_\infty(P)<\theta(\sysp)$. The following example is provided.
\begin{exmp}
 Let $\tau=0.4$ and $P(s)= \frac{(s+5)(s^2 + 3s + 102.3)}{(s+1)(s^2 + 6s + 109)}$.
One can verify that $\theta_\infty(P)$ in \eqref{eq:siso_LTI_sphase} is attained at $\omega_0=2.613$ and thus $a_0=1.3092, b_0=-1.332$. Let $\omega_1=10$ so that $a_1=0.5286, b_1=-0.1577$. Clearly, $\abs{\angle z_0} \geq \abs{\angle z_1}$. Then, we have
 \begin{align*}
 \cos\theta(\sysp) &\leq \textstyle \frac{0.4\times1.3092+0.6\times0.5286}{\sqrt{0.4\times3.4882+0.6\times0.3043
}} =0.6694,\\
 \cos\theta_\infty(P)&=\textstyle\frac{1.3092}{\sqrt{3.4882}}=0.701.
 \end{align*}
 This implies $\cos\theta(\sysp)<\cos\theta_\infty(P)$ and ${\theta_\infty}(P)< \theta(\sysp)$.
 \end{exmp}
 
{\bfseries PROOF of Corollary~\ref{coro:lure}.}
First, $\theta(\sysn)\leq \arccos \frac{2\sqrt{ab}}{a+b}$ $<\frac{\pi}{2}$ by Corollary~\ref{prop:nonlinearity}. By hypothesis, when $\angle P(\jw)>\frac{\pi}{2}$ or $\angle P(\jw)<-\frac{\pi}{2}$ holds for some $\omega\in\interval[open]{\infty}{\infty}$, it holds that $\theta_\infty(P)$ in \eqref{eq:siso_LTI_sphase} is contained in $\interval[open]{\frac{\pi}{2}}{\pi-\arccos \frac{2\sqrt{ab}}{a+b}}$. Then, for $\tau>0$, we have
 $\theta(\tau \sysp)+\theta(\sysn)=\theta(\sysp)+\theta(\sysn)=\theta_\infty(P)+\theta(
 \sysn)<\pi$,
 where the second equality uses that $\theta_\infty(P)=\theta(\sysp)$ by invoking Proposition~\ref{prop:SISO}. In this case, by Theorem~\ref{thm:spt}, $\rbkt{\tau {\sysp}}\,\#\,{\sysn}$ is stable for all $\tau >0$. When $\abs{\angle P(\jw)}\leq \frac{\pi}{2}$ holds for all $\omega\in\interval{-\infty}{\infty}$, $\tau \sysp$ is passive, and $\sysn$ is very strictly passive, as stated before. In this case, by the passivity theorem \cite[Sec.~6.6]{Vidyasagar:93}, $\rbkt{\tau {\sysp}}\,\#\,{\sysn}$ is stable for all $\tau >0$. \hfill $\square$

\section*{Acknowledgment}
The authors would like to thank Li Qiu and Wei Chen for useful discussions and helpful comments. 
 
\bibliographystyle{abbrv}
\bibliography{SingularAngle}

\begin{thebibliography}{10}

\bibitem{Anderson:88}
B.~D.~O. Anderson and M.~Green.
\newblock Hilbert transform and gain/phase error bounds for rational functions.
\newblock {\em IEEE Trans. Circuits Syst.}, 35(5):528--535, 1988.

\bibitem{Anderson:73}
B.~D.~O. Anderson and S.~Vongpanitlerd.
\newblock {\em Network Analysis and Synthesis: {A} Modern Systems Theory Approach}.
\newblock Prentice-Hall, Englewood Cliffs, NJ, 1973.

\bibitem{Angeli:06}
D.~Angeli.
\newblock Systems with counterclockwise input-output dynamics.
\newblock {\em IEEE Trans. Autom. Control}, 51(7):1130--1143, 2006.

\bibitem{Arcak:06}
M.~Arcak and E.~D. Sontag.
\newblock Diagonal stability of a class of cyclic systems and its connection with the secant criterion.
\newblock {\em Automatica}, 42(9):1531--1537, 2006.

\bibitem{Astrom:10}
K.~J. {\AA}str{\"o}m and R.~M. Murray.
\newblock {\em Feedback Systems: {A}n Introduction for Scientists and Engineers}.
\newblock Princeton University Press, Princeton, NJ, 2010.

\bibitem{Auzinger:03}
W.~Auzinger.
\newblock Sectorial operators and normalized numerical range.
\newblock {\em Appl. Numer. Math.}, 45(4):367--388, 2003.

\bibitem{bar-on:90}
J.~R. Bar-on and E.~A. Jonckheere.
\newblock Phase margins for multivariable control systems.
\newblock {\em Int. J. Control}, 52(2):485--498, 1990.

\bibitem{Billings:94}
S.~A. Billings and H.~Zhang.
\newblock Analysing non-linear systems in the frequency domain--{\uppercase\expandafter{\romannumeral2}}. {T}he phase response.
\newblock {\em Mech. Syst. Signal Process.}, 8(1):45--62, 1994.

\bibitem{Carrasco:16}
J.~Carrasco, M.~C. Turner, and W.~P. Heath.
\newblock {Z}ames--{F}alb multipliers for absolute stability: {F}rom {O}'{S}hea's contribution to convex searches.
\newblock {\em Eur. J. Control}, 28:1--19, 2016.

\bibitem{Chaffey:21c}
T.~Chaffey, F.~Forni, and R.~Sepulchre.
\newblock Scaled relative graphs for system analysis.
\newblock In {\em Proc. 60th IEEE Conf. Decision and Control}, pages 3166--3172, Austin, TX, 2021.

\bibitem{Chaffey:21j}
T.~Chaffey, F.~Forni, and R.~Sepulchre.
\newblock Graphical nonlinear system analysis.
\newblock {\em IEEE Trans. Autom. Control}, 68(10):6067--6081, 2023.

\bibitem{Chen:20j}
C.~Chen, D.~Zhao, W.~Chen, S.~Z. Khong, and L.~Qiu.
\newblock Phase of nonlinear systems.
\newblock {\em arXiv preprint arXiv:2012.00692}, 2021.

\bibitem{Chen:20c}
C.~Chen, D.~Zhao, W.~Chen, and L.~Qiu.
\newblock A nonlinear small phase theorem.
\newblock In {\em Late Breaking Results of 21st IFAC World Congress}, Berlin, Germany, 2020.

\bibitem{Chen:98}
J.~Chen.
\newblock Multivariable gain-phase and sensitivity integral relations and design trade-offs.
\newblock {\em IEEE Trans. Autom. Control}, 43(3):373--385, 1998.

\bibitem{Chen:19}
W.~Chen, D.~Wang, S.~Z. Khong, and L.~Qiu.
\newblock Phase analysis of {MIMO} {LTI} systems.
\newblock In {\em Proc. 58th IEEE Conf. Decision and Control}, pages 6062--6067, Nice, France, 2019.

\bibitem{Chen:21}
W.~Chen, D.~Wang, S.~Z. Khong, and L.~Qiu.
\newblock A phase theory of multi-input multi-output linear time-invariant systems.
\newblock {\em SIAM J. Control Optim.}, 62(2):1235--1260, 2024.

\bibitem{Chua:79}
L.~O. Chua and C.-Y. Ng.
\newblock Frequency domain analysis of nonlinear systems: {G}eneral theory.
\newblock {\em IEE J. Electr. Circuits Syst.}, 3(4):165--185, 1979.

\bibitem{Coutin:10}
D.~F. Coutinho, M.~Fu, and C.~E. de~Souza.
\newblock Input and output quantized feedback linear systems.
\newblock {\em IEEE Trans. Autom. Control}, 55(3):761--766, 2010.

\bibitem{Freudenberg:88}
J.~S. Freudenberg and D.~P. Looze.
\newblock {\em Frequency Domain Properties of Scalar and Multivariable Feedback Systems}.
\newblock Springer, Berlin, Germany, 1988.

\bibitem{Fu:05_2}
M.~Fu and L.~Xie.
\newblock The sector bound approach to quantized feedback control.
\newblock {\em IEEE Trans. Autom. Control}, 50(11):1698--1711, 2005.

\bibitem{Ghallab:22}
A.~G. Ghallab and I.~R. Petersen.
\newblock Negative imaginary systems theory for nonlinear systems: {A} dissipativity approach.
\newblock {\em arXiv preprint arXiv:2201.00144}, 2022.

\bibitem{Gustafson:94}
K.~Gustafson.
\newblock Antieigenvalues.
\newblock {\em Linear Algebra Appl.}, 208:437--454, 1994.

\bibitem{Gustafson:97}
K.~E. Gustafson and D.~K. Rao.
\newblock {\em Numerical Range: The Field of Values of Linear Operators and Matrices}.
\newblock Springer, New York, NY, 1997.

\bibitem{Hill:91}
D.~J. Hill.
\newblock A generalization of the small-gain theorem for nonlinear feedback systems.
\newblock {\em Automatica}, 27(6):1043--1045, 1991.

\bibitem{Hill:80}
D.~J. Hill and P.~J. Moylan.
\newblock Dissipative dynamical systems: {B}asic input-output and state properties.
\newblock {\em J. Franklin Inst.}, 309(5):327--357, 1980.

\bibitem{Jiang:18}
Z.-P. Jiang and T.~Liu.
\newblock Small-gain theory for stability and control of dynamical networks: A survey.
\newblock {\em Annu. Rev. Control.}, 46:58--79, 2018.

\bibitem{Jiang:94}
Z.~P. Jiang, A.~R. Teel, and L.~Praly.
\newblock Small-gain theorem for {ISS} systems and applications.
\newblock {\em Math. Control. Signals, Syst.}, 7:95--120, 1994.

\bibitem{Jonsson:01}
U.~J{\"o}nsson.
\newblock Lecture notes on integral quadratic constraints.
\newblock Department of Mathematics, KTH Royal Institute of Technology, 2001.

\bibitem{Khalil:02}
H.~K. Khalil.
\newblock {\em Nonlinear Systems}.
\newblock Prentice Hall, Upper Saddle River, NJ, 3rd edition, 2002.

\bibitem{Khong:21}
S.~Z. Khong.
\newblock On integral quadratic constraints.
\newblock {\em IEEE Trans. Autom. Control}, 67(3):1603--1608, 2022.

\bibitem{Khong:22}
S.~Z. Khong, C.~Chen, and A.~Lanzon.
\newblock Feedback stability analysis via dissipativity with dynamic supply rates.
\newblock {\em Automatica}, page 112000, 2025.

\bibitem{Khong:18}
S.~Z. Khong, I.~R. Petersen, and A.~Rantzer.
\newblock Robust stability conditions for feedback interconnections of distributed-parameter negative imaginary systems.
\newblock {\em Automatica}, 90:310--316, 2018.

\bibitem{Krein:69}
M.~G. Krein.
\newblock Angular localization of the spectrum of a multiplicative integral in a {H}ilbert space.
\newblock {\em Funct. Anal. Appl.}, 3(1):73--74, 1969.

\bibitem{Lanzon:22}
A.~Lanzon and P.~Bhowmick.
\newblock Characterization of input--output negative imaginary systems in a dissipative framework.
\newblock {\em IEEE Trans. Autom. Control}, 68(2):959--974, 2023.

\bibitem{Lanzon:17}
A.~Lanzon and H.-J. Chen.
\newblock Feedback stability of negative imaginary systems.
\newblock {\em IEEE Trans. Autom. Control}, 62(11):5620--5633, 2017.

\bibitem{Megretski:97}
A.~Megretski and A.~Rantzer.
\newblock System analysis via integral quadratic constraints.
\newblock {\em IEEE Trans. Autom. Control}, 42(6):819--830, 1997.

\bibitem{Owens:84}
D.~Owens.
\newblock The numerical range: {A} tool for robust stability studies?
\newblock {\em Syst. Control Lett.}, 5(3):153--158, 1984.

\bibitem{Pates:21}
R.~Pates.
\newblock The scaled relative graph of a linear operator.
\newblock {\em arXiv preprint arXiv:2106.05650}, 2021.

\bibitem{Petersen:10}
I.~R. Petersen and A.~Lanzon.
\newblock Feedback control of negative-imaginary systems.
\newblock {\em IEEE Control Systems Magazine}, 30(5):54--72, 2010.

\bibitem{Postlethwaite:81}
I.~Postlethwaite, J.~Edmunds, and A.~MacFarlane.
\newblock Principal gains and principal phases in the analysis of linear multivariable feedback systems.
\newblock {\em IEEE Trans. Autom. Control}, 26(1):32--46, 1981.

\bibitem{Rantzer:19}
A.~Rantzer.
\newblock Lecture notes on nonlinear control and servo systems, {L}und University, 2019.

\bibitem{Rantzer:97}
A.~Rantzer and A.~Megretski.
\newblock {\em System Analysis via {I}ntegral {Q}uadratic {C}onstraints {P}art~{\uppercase\expandafter{\romannumeral2}}: {A}bstract Theory}.
\newblock Technical Reports TFRT-7559. Lund Institute of Technology, 1997.

\bibitem{Rugh:81}
W.~J. Rugh.
\newblock {\em Nonlinear System Theory: The Volterra/Wiener Approach}.
\newblock The Johns Hopkins University Press, Baltimore, MD, 1981.

\bibitem{Ryu:21}
E.~K. Ryu, R.~Hannah, and W.~Yin.
\newblock Scaled relative graphs: {N}onexpansive operators via {2D} {E}uclidean geometry.
\newblock {\em Math. Program.}, pages 1--51, 2021.

\bibitem{Sepulchre:97}
R.~Sepulchre, M.~Jankovi\'{c}, and P.~V. Kokotovi\'{c}.
\newblock {\em Constructive Nonlinear Control}.
\newblock Springer-Verlag London, London, UK, 1997.

\bibitem{Sontag:06}
E.~D. Sontag.
\newblock Passivity gains and the ``secant condition'' for stability.
\newblock {\em Syst. Control Lett.}, 55(3):177--183, 2006.

\bibitem{Stancu:12}
I.~M. Stancu-Minasian.
\newblock {\em Fractional Programming: {T}heory, Methods and Applications}.
\newblock Springer Dordrecht, Dordrecht, Netherlands, 1997.

\bibitem{Tits:99}
A.~L. Tits, V.~Balakrishnan, and L.~Lee.
\newblock Robustness under bounded uncertainty with phase information.
\newblock {\em IEEE Trans. Autom. Control}, 44(1):50--65, 1999.

\bibitem{Van:17}
A.~van~der Schaft.
\newblock {\em $\mathcal{L}_2$-Gain and Passivity Techniques in Nonlinear Control}.
\newblock Springer International Publishing AG, Cham, Switzerland, 3rd edition, 2017.

\bibitem{Vidyasagar:77}
M.~Vidyasagar.
\newblock $\mathcal{L}_2$-stability of interconnected systems using a reformulation of the passivity theorem.
\newblock {\em IEEE Trans. Circuits Syst.}, 24(11):637--645, 1977.

\bibitem{Vidyasagar:93}
M.~Vidyasagar.
\newblock {\em Nonlinear Systems Analysis}.
\newblock Prentice-Hall, Englewood Cliffs, NJ, 2nd edition, 1993.

\bibitem{Wielandt:67}
H.~Wielandt.
\newblock {\em Topics in the Analytic Theory of Matrices}.
\newblock University of Wisconsin Lecture Notes, Madison, WI, 1967.

\bibitem{Zames:66}
G.~Zames.
\newblock On the input-output stability of time-varying nonlinear feedback systems {P}art {\uppercase\expandafter{\romannumeral1}}: Conditions derived using concepts of loop gain, conicity, and positivity.
\newblock {\em IEEE Trans. Autom. Control}, 11(2):228--238, 1966.

\bibitem{Zames:66_2}
G.~Zames.
\newblock On the input-output stability of time-varying nonlinear feedback systems {P}art {\uppercase\expandafter{\romannumeral2}}: Conditions involving circles in the frequency plane and sector nonlinearities.
\newblock {\em IEEE Trans. Autom. Control}, 11(3):465--476, 1966.

\bibitem{Zames:68}
G.~Zames and P.~L. Falb.
\newblock Stability conditions for systems with monotone and slope-restricted nonlinearities.
\newblock {\em SIAM J. Control}, 6(1):89--108, 1968.

\bibitem{Zhao:23_Angle}
D.~Zhao, C.~Chen, and J.~Chen.
\newblock Small gain and small angle conditions for feedback stability analysis of linear stochastic systems.
\newblock {\em IEEE Trans. Autom. Control}, 69(5):3349--3356, 2024.

\bibitem{Zhao:22_NI}
D.~Zhao, C.~Chen, and S.~Z. Khong.
\newblock A frequency-domain approach to nonlinear negative imaginary systems analysis.
\newblock {\em Automatica}, 146:110604, 2022.

\bibitem{Zhou:96}
K.~Zhou, J.~Doyle, and K.~Glover.
\newblock {\em Robust and Optimal Control}.
\newblock Prentice Hall, Englewood Cliffs, NJ, 1996.

\end{thebibliography}

\end{document}